\documentclass[
reprint,
superscriptaddress,
amsmath,
amssymb,
longbibliography,twocolumn
]{revtex4-2}

\usepackage{amsmath}
\usepackage{mathtools}
\usepackage{amssymb}
\usepackage{graphicx}
\usepackage{bm}
\usepackage[colorlinks, linkcolor=blue, citecolor=blue, urlcolor=blue,breaklinks=true]{hyperref}
\usepackage{color,dsfont,multirow,booktabs}
\usepackage{braket}
\usepackage{tabularx}
\usepackage{lipsum,booktabs}

\begin{document}
\preprint{APS/123-QED}
\title{Enhanced quantum emission from a topological Floquet resonance}

\author{Shirin Afzal}
\affiliation{Institute for Quantum Science and Technology, and Department of Physics and Astronomy University of Calgary,
2500 University Drive NW, Calgary, Alberta T2N 1N4, Canada}
\author{Tyler J. Zimmerling}
\affiliation{Department of Electrical and Computer Engineering, University of Alberta, 9211 116 Street NW, Edmonton, Alberta, Canada  T6G 1H9}
\author{Mahdi Rizvandi}
\affiliation{Institute for Quantum Science and Technology, and Department of Physics and Astronomy University of Calgary,
2500 University Drive NW, Calgary, Alberta T2N 1N4, Canada}
\author{Majid Taghavi}
\affiliation{Institute for Quantum Science and Technology, and Department of Physics and Astronomy University of Calgary,
2500 University Drive NW, Calgary, Alberta T2N 1N4, Canada}
\author{Leili Esmaeilifar}
\affiliation{Institute for Quantum Science and Technology, and Department of Physics and Astronomy University of Calgary,
2500 University Drive NW, Calgary, Alberta T2N 1N4, Canada}
\author{Taras Hrushevskyi }
\affiliation{Institute for Quantum Science and Technology, and Department of Physics and Astronomy University of Calgary,
2500 University Drive NW, Calgary, Alberta T2N 1N4, Canada}
\author{Manpreet Kaur}
\affiliation{Institute for Quantum Science and Technology, and Department of Physics and Astronomy University of Calgary,
2500 University Drive NW, Calgary, Alberta T2N 1N4, Canada}
\author{Vien Van}
\affiliation{Department of Electrical and Computer Engineering, University of Alberta, 9211 116 Street NW, Edmonton, Alberta, Canada  T6G 1H9}
\author{Shabir Barzanjeh}
\email{Corresponding Author: shabir.barzanjeh@ucalgary.ca}
\affiliation{Institute for Quantum Science and Technology, and Department of Physics and Astronomy University of Calgary,
2500 University Drive NW, Calgary, Alberta T2N 1N4, Canada}


\begin{abstract}
Entanglement is a valuable resource in quantum information technologies. The practical implementation of entangled photon sources faces obstacles from imperfections and defects inherent in physical systems, resulting in a loss or degradation of entanglement. The topological photonic insulators, however, have emerged as promising candidates, demonstrating an exceptional capability to resist defect-induced scattering, thus enabling the development of robust entangled sources. Despite their inherent advantages, building programmable topologically protected entangled sources remains challenging due to complex device designs and weak material nonlinearity. Here we present a development in entangled photon pair generation achieved through a non-magnetic and tunable anomalous Floquet insulator, utilizing an optical spontaneous four-wave mixing process. We verify the non-classicality and time-energy entanglement of the photons generated by our topological system. Our experiment demonstrates a substantial enhancement in nonclassical photon pair generation compared to devices reliant only on topological edge states. Our result could lead to the development of resilient quantum sources with potential applications in quantum technology.
\end{abstract}
\maketitle

\begin{figure*}
    \centering
 \includegraphics [width=0.9\linewidth]{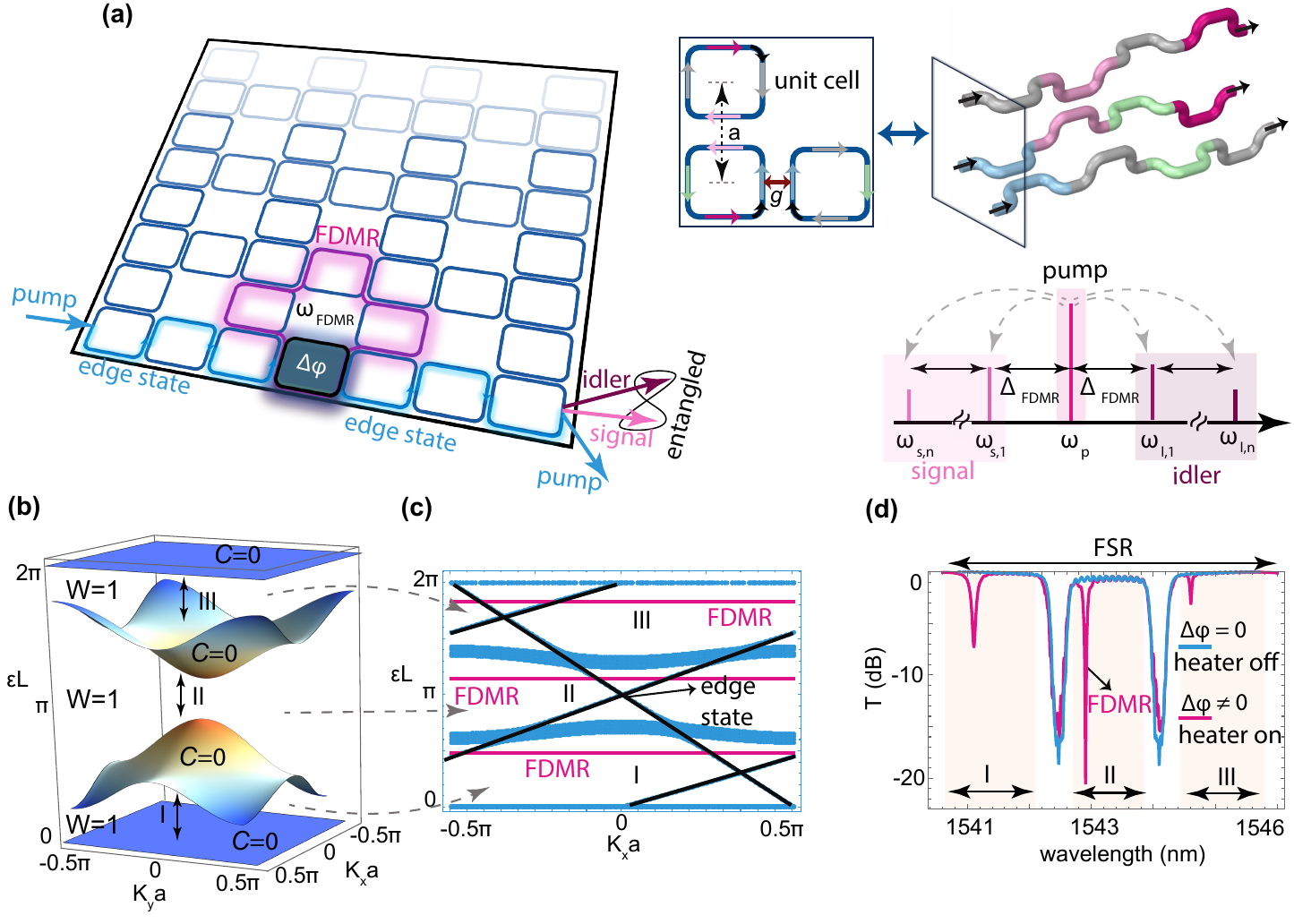}
    \caption{\textbf{Schematic representation and simulation of a Floquet topological entanglement source.} \textbf{a,} A 2D square microring lattice consisting of individual ring resonators arranged in identical unit cells. Each unit cell (inset) comprises three closely coupled microrings arranged in a square formation with the coupling gap $g$ and center-to-center distance $a$. By unrolling each microring from the points shown by black arrows, an equivalent coupled waveguide array can be obtained (inset). Four periodic coupling sequences, in the microring unit cell and its equivalent waveguide array, are shown by light blue, light pink, light green, and dark pink arrows/segments. The gray color shows the uncoupled part of microrings/waveguides. An FDMR with resonance frequency $\omega_{\mathrm{FDMR}}$ can be created by utilizing a thermo-optics effect to introduce a phase shift $\Delta \phi$ in one of the rings along the path of the edge state. The lattice's periodicity leads to the appearance of multiple equally spaced modes, separated by the frequency of $\Delta_\mathrm{FDMR}$. By applying a pump to the system via the edge state and utilizing the SFWM process within the FDMR, we can produce idler-signal entangled photons at various $\Delta_\mathrm{FDMR}$ within the system, as illustrated in the lower inset.  \textbf{b,} The computed band structure for the photonic anomalous Floquet topological insulator in one Floquet–Brillouin zone with wavenumbers $K_x$ and $K_y$, 
     comprising one unit cell periodic in both $x$ and $y$ directions. The transmission bands of Floquet states, with Chern number $C=0$, are separated by three topological bandgaps, labeled by (I, II, II),  with winding number $W=1$. \textbf{c,} Projected band diagrams of a semi-infinite lattice limited in the $y$-direction (consisting of 10 unit cells) and extending infinitely in the $x$-direction. The black lines show two pseudospin topological edge states in each bandgap. The pink bands represent the flat bands of Floquet bulk, obtained using a periodic $5 \times 5$ supercell, which are distinct from the transmission band manifolds due to a phase detune. \textbf{d,} Simulated transmission spectra of an ideal topological photonic insulator lattice of identical microring resonators with power coupling of $98\%$, considering intrinsic loss, in the presence (pink line) and absence (blue line) of the phase shift $\Delta \phi$.}
    \label{Fig1}
\end{figure*}
\section{INTRODUCTION}
The unique properties of entanglement and topology have enormous potential for advancing quantum technology.  Entanglement has a wide range of applications in quantum information \cite{RevModPhys.81.865,gisin2014quantum}, quantum sensing \cite{Aslam2023, Pirandola2018}, computing \cite{Ladd2010, Harrow2017},  and communication \cite{couteau2023applications, Hanson}. Significant progress has been made in the optical domain to generate entangled states using integrated devices \cite{Wang2020, Pelucchi2022}, opening up avenues for practical real-world applications \cite{kues2017chip, shadbolt2012generating, matthews2011heralding, Lu2019}. However, preserving and effectively harnessing entanglement face challenges due to inevitable fabrication imperfections and defects in the microchips \cite{Topolancik, Luca}. These disorders can lead to loss, scattering, and spectral randomness, ultimately affecting the stability and purity of the entanglement \cite{Metcalf2014, Burgwal:17}.

Similarly, topological photonic insulators (TPIs) have attracted considerable attention due to their ability to guide light through topologically-protected edge states \cite{lu2014topological, Khanikaev2017, ozawa2019topological}. These states have applications in different domains such as lasing \cite{Bandres, st2017lasing, bahari2017nonreciprocal}, soliton generation \cite{Kirsch2021, mukherjee2020observation} and frequency comb \cite{Mittal2021}. The most important aspect of TPIs is their resilience to scattering caused by defects, making them highly desirable for applications in photon transport \cite{rechtsman2016topological, Barik,segev2020topological}. The TPI systems can also be utilized for generating nonclassical emissions, offering exceptional possibilities for advancing quantum technology.

Recently, TPIs have been used for generating entangled photons using coupled waveguides \cite{Redondo} and static microring resonators \cite{Mittal2018, Mittal2021HOM}, as well as through Floquet-based TPIs strongly coupled microring resonators \cite{Dai2022}. Nevertheless, the practical implementation of nonclassical sources based on edge states has faced challenges due to the weak nonlinear properties of the materials. Ensuring sufficient pair generation requires either long lattice boundaries, leading to increased material loss, or utilizing high pump powers that introduce noise and reduce the purity of the entanglement. These limitations have constrained the potential applications of TPIs in quantum technologies, especially in scenarios demanding bright entangled emissions, such as long-distance quantum communications \cite{PhysRevLett.93.180502, Yu2020}. Despite demonstrating resonance enhancement of photon pair generation through spontaneous four-wave mixing (SFWM) in conventional resonators \cite{Kumar:13, ma2017silicon,ma2020ultrabright, afifi2021contra,grassani2015micrometer, wakabayashi2015time, guo2018generation, clementi2023programmable}, the experimental realization of a bright Floquet TPI entanglement source has remained elusive and unexplored until now.

Here we explore entanglement generation in a novel resonance effect called Floquet Defect Mode Resonance (FDMR) in the bulk of a Floquet TPI \cite{Afzal2021}. We present experimental evidence that demonstrates a substantial enhancement in entangled photon pair generation by harnessing this compact topological cavityless resonance effect. By exciting the wavelength-tunable FDMR and utilizing SFWM in a silicon microring lattice coupled to a topologically protected edge state, we create a bright quantum photon source. This source demonstrates a second-order cross-correlation of photon pairs approximately 3300 times higher compared to using edge mode without resonance with two-photon interference visibility of $98.2\%$. These results demonstrate the potential of FDMR in significantly improving the efficiency of entangled photon pair generation in TPI-based systems, thereby removing the need for high power or extended chip boundaries. Our topological quantum device combines the advantages of integrated topological photonics, tunable localized flat-band optical modes, and robustness to disorder, representing a novel approach to generating entangled photon pairs. Additionally, FDMRs can be selectively turned on and off and excited anywhere in a 2D topological system, a capability not easily achieved in conventional non-topological systems. This allows for the creation of multiple identical quantum sources, where both the wavelength and bandwidth of the generated entangled photons are similar. By selectively deactivating the resonator affected by fabrication imperfections and activating another FDMR away from the defect, we can access a high-quality resonance mode required to enhance the generation of entangled photon pairs. We note that compared to topologically trivial systems such as single rings, our topological system may not offer performance advantages. However, our topological photonic circuit provides an alternative method for practical situations where ring resonators may be ineffective. For example, in complex quantum circuits with multiple photon pair sources, our Floquet lattice with multiple FDMRs can offer robustness due to topological protection, unlike multiple separate single rings. This work can potentially lead to the development of TPI high-dimensional entangled quantum states \cite{Kues2017} as well as quantum logic \cite{Imany2019}.
\section{Theory: Trapping light in a topological Floquet insulator}
Figure \ref{Fig1}a presents the schematic of our Floquet topological photonic insulator, created using two-dimensional (2D) directly-coupled microring resonators \cite{zimmerling2022broadband}.
Each unit cell, see the inset of Fig. \ref{Fig1}a, comprises three strongly coupled identical microrings arranged in a square shape with
the coupling coefficient $p_c$.
As light propagates around each microring, it evanescently couples to the neighboring rings in a periodic sequence, see Fig. 1(a), with the period equal to the microring circumference $L$. The system thus emulates a periodically driven system with the evolution along the direction of light propagation $z$ rather than time \cite{Afzal2018}. By varying the coupling between microrings, the topological phase of the lattice can be tuned, leading to the appearance of Chern or anomalous Floquet topological insulators respectively in the weak or strong coupling regimes,  as shown in Refs \cite{Liang2013, Afzal2018,afzal2020realization, Dai2022}. Our 2D microring lattice follows a similar framework as the Floquet systems and satisfies the following eigenvalue equation for the wavefunction $|\psi_n(\textbf{k})\rangle$

\begin{equation}
    U_F(\textbf k)|\psi_n(\textbf k)\rangle=e^{i\epsilon_n(\textbf k) L}|\psi_n(\textbf k)\rangle,
    \label{wavefunction}
\end{equation}
where $\epsilon_n(\textbf k)$ is the quasi-energy band of the lattice with the periodicity of $2\pi/L$. The Floquet operator $U_F(\textbf k)=\mathcal{T}e^{i\int_0^L H(\textbf k, z')dz'}$, where $\mathcal{T}$ represents the time-order operator, depends on the Floquet-Bloch Hamiltonian $H(\textbf k, z)=H_{\mathrm{FB}}(\textbf k, z)$ that exhibits periodicity along the $z$ direction $H(\textbf k, z)=H(\textbf k, z+L)$, with a period of $L$. This characteristic mimics the behavior of a periodically driven Hamiltonian, where the variable $z$ plays the role of time $z\rightarrow{ t}$. We can obtain an expression for $H(\textbf k, z)$ by transforming the microring lattice into an equivalent coupled waveguide array, see the inset of Fig. \ref{Fig1}a.  In the strong coupling regime, our 2D lattice exhibits three bandgaps with non-zero winding numbers. All bands possess trivial Chern numbers $C=0$, making it an anomalous Floquet insulator, as shown in the band structure of one Floquet-Brillouin zone of the unit cell in Fig. \ref{Fig1}b. To verify the presence of topological edge states, we impose boundaries along the $y$-axis while assuming the lattice extends infinitely along the $x$-axis. From the projected quasi-energy band diagram, we observe the existence of
two pseudospin 
topological edge states in each bandgap \cite{afzal2020realization}, as shown in Fig. \ref{Fig1}c by black traces.

By exploiting the natural hopping sequence of our 2D lattice,
we can achieve the confinement of light within a closed loop, creating a cavityless local resonator. This confinement is attained by introducing a small perturbation in the driving sequence, in the form of a phase shift ($\Delta \phi$), in one of the ring resonators along the edge state's trajectory, leading to the formation of a flat-band Floquet mode within the 2D lattice. As a result, the light becomes effectively trapped within the loop, leading to a locally confined mode referred to as FDMR. This perturbation subsequently modifies the total Hamiltonian of the system $H=H_{\mathrm{FB}}+H_\mathrm{FDMR}$ where $
    H_\mathrm{FDMR}=\hbar \omega_\mathrm{FDMR} a^\dagger a$ 
describes the FMDR with a resonance frequency $\omega_\mathrm{FDMR}$ and annihilation (creation) operator $a$ ($a^\dagger$). The associated quasienergy of this mode experiences a shift directly proportional to the magnitude of the induced phase, as depicted in Fig. \ref{Fig1}c by pink lines (see Supplementary Information). This FDMR mode can achieve a large Q-factor, approaching~$10^5$, compared to other 2D topological resonators \cite{gao2020dirac, shao2020high}, mainly because it lacks physical boundaries that would otherwise confine light within the defect resonator. The resonance pattern in FDMR follows the trajectory of the topologically nontrivial Floquet bulk mode with a total circumference of $3L$. Figure \ref{Fig1}d illustrates the simulation of the transmission spectrum of our design in the presence and absence of a phase shift, demonstrating the appearance of the FDMR. Note that, this mode is coupled to the edge state existing within the same bandgap, implying that we can control or access the FDMR via the edge state. 

To generate correlated photon pairs through the SFWM process in the bandgap of the topological 2D lattice, we utilize the third-order nonlinearity $\chi^3$ in silicon. The Hamiltonian that describes this process is given by
\begin{equation}
    H_\mathrm{non}=\hbar g_\text{nl} \big(a_p a_p a_s^\dagger a_i^\dagger+h.c\big),
    \label{Hnonlinear}
\end{equation}
where $g_\text{nl}$ is the strength of the SFWM while $a_p$ and $a_{s(i)}$ refer to the annihilation operators of the pump and signal (idler) modes, respectively. The Hamiltonian represents a four-photon mixing process in which two photons from the pump are annihilated, resulting in the creation of photons in the idler and signal modes, which were initially in a vacuum state. With the presence of the FDMR, the generation of the photon pairs is significantly enhanced, surpassing the pair generation rate achieved solely with the edge state
\begin{equation}
    N_\mathrm{FDMR}\propto N_\mathrm{edge} \times  Q^3,
    \label{N_FDMR_vs_Q}
\end{equation}
where $N_\mathrm{FDMR}$ and $N_\mathrm{edge}$ are the pair generation rates of the FDMR and edge state, respectively, and $Q$ is the quality factor of the FDMR, see the Supplementary Information.



\begin{figure*}
    \centering
 \includegraphics [width=0.9\linewidth]{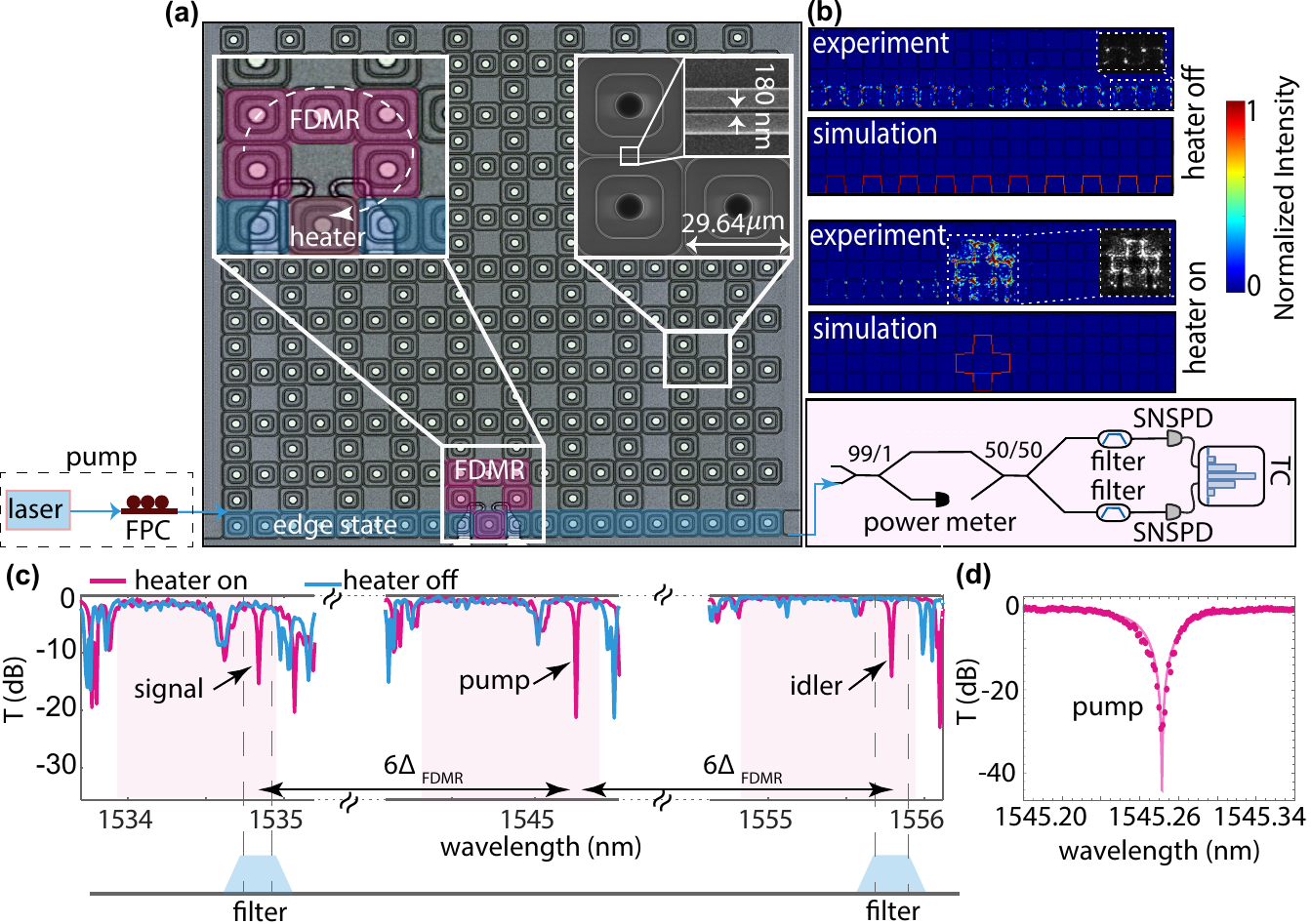}
    \caption{\textbf{Experimental realization.} \textbf{a,} 
    A microscopic image of the anomalous Floquet topological insulator lattice, featuring a $10\times 10$ unit cells arrangement across a $595.32\times 595.32$ $\mu$m chip. The insets provide detailed scanning electron microscopy (SEM) views of a unit cell and the structure of the heater functioning as a thermo-optics phase shifter. Each microring in our design has a square shape, with side lengths of 29.64 $\mu$m, and rounded corners with a radius of 5 $\mu$m to minimize scattering losses. 
     Additionally, the figure includes a diagram of the experimental setup, which involves a pump laser passing through a fiber polarization controller (FPC) and being injected into the sample via a lensed fiber. The transmitted light is measured using a photon detector and a power meter, two narrow bandpass filters, two superconducting nanowire single-photon detectors (SNSPDs), and a time controller (TC). \textbf{b,} Near Infrared images of the edge state and FDMR in real-time and comparison with analytical simulations. \textbf{c,} The normalized transmission (T) of the lattice as a function of the pump wavelength. \textbf{d,} Zoomed-in transmission of the FDMR with respect to the wavelength of the pump. The solid line represents the fitted theoretical model, as detailed in the Supplementary Information.}
    \label{Fig2}
\end{figure*}

\begin{figure*}
    \centering
 \includegraphics [width=0.9\linewidth]{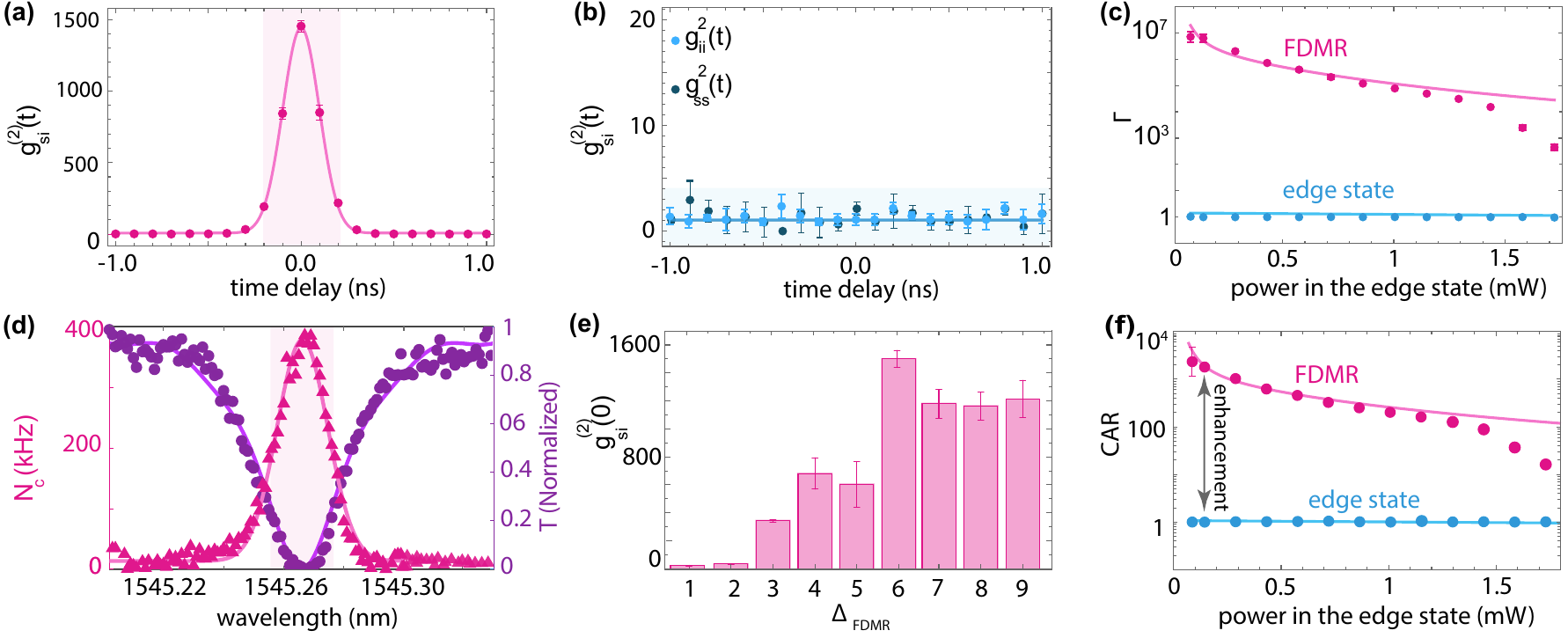}
    \caption{\textbf{Enhancement of the photon pairs generation}.  \textbf{a} and \textbf{b,} show the measured second-order cross-correlation $g_{si}^{(2)}(t)$ and the second-order autocorrelation function $g_{ss/ii}^{(2)}(t)$, respectively, for the power of 0.29 mW in the edge state. \textbf{c,} Obtained nonclassical parameter, $\Gamma$, for both FDMR and edge state at different pump powers. Increasing the pump power pushes the system to a nonlinear regime such as the two-photon absorption (TPA). 
    \textbf{d,} Illustrates the transmission, purple circles, and coincidence rate, pick triangles, (after subtracting the noise counts) versus pump wavelengths around the resonance frequency of the FDMR for a pump power of 0.29 mW. The measured bandwidth of the coincidence rate (highlighted region) is 21 pm, which is approximately $75\%$ of the bandwidth of the FDMR (28 pm). \textbf{e,} Shows the measured $g_{si}^{(2)}(0)$ at different $\Delta_\mathrm{FDMR}$ at the same power. \textbf{f,} Display the obtained  CAR for FDMR and edge state versus pump power in the edge state. For all panels, the error bars are derived from three independent measurements. For some points, the size of the dots is larger than the error bars. Fitted solid lines in panels \textbf{a} and \textbf{b} utilize the Poissonian distribution while fitting for panels \textbf{c}, \textbf{d}, and \textbf{f} is based on the comprehensive theory outlined in the Supplementary Information.}
    \label{Fig3}
\end{figure*}

\section{Sample Characterization and Measuring FDMR}
We experimentally demonstrate the generation and enhancement of entangled photons within our 2D Floquet lattice. The lattice is structured with a $10\times 10$ unit cell arrangement and fabricated on a Silicon-on-Insulator (SOI) substrate, incorporating a total of 300 individual ring resonators, see Fig. \ref{Fig2}a. 
Each microring in our design is square-shaped and is designed
to achieve efficient power coupling, enabling around
$p_c^2=98\%$ of the power transfer from one microring to the nearest microring. This design is specifically chosen to achieve the desired anomalous Floquet insulator behavior around the wavelength of 1545 nm. 

Figure \ref{Fig2}b presents the real-time field distribution for the edge state and FDMR, as measured via a Near-Infrared camera. These experimental results demonstrate good agreement with the analytical calculations obtained by solving the Schrödinger equation. In particular, 
the existence of the edge state provides evidence of 
distinct nontrivial light propagation along the boundaries of the lattice. By applying a phase shift in a microring located on the bottom edge of the sample, we observe the emergence of the FDMR mode, characterized by a strongly localized field distribution in a loop pattern.

The chip's transmission spectrum is measured by injecting a laser into the lattice at the bottom left edge and detecting the output light at the bottom right edge using a power meter, as displayed in Fig. \ref{Fig2}a. During the measurement, the laser's wavelength is swept from 1530 nm to 1557 nm, equivalent to 13 $\Delta_{\mathrm{FDMR}}$. 
Note that we define $\Delta_{\mathrm{FDMR}}$ as the frequency spacing of the FDMR which is approximately one-third of the FSR of our TPI, measuring 1.72 nm. The normalized transmission spectrum, shown in blue in Fig. \ref{Fig2}c, allows us to identify topologically nontrivial bandgaps, characterized by regions of high and flat transmission, highlighted with light pink. Upon activation of the heater (phase shift), a Floquet bulk mode is lifted into the bandgap, resulting in a flattened energy band and a resonant mode spatially localized in a loop pattern, as demonstrated at the bottom of Fig. \ref{Fig2}a (see the inset). The transmission spectrum in the output waveguide, with a phase detune of $\Delta \phi=2.37\pi$, is represented by the pink trace in Fig. \ref{Fig2}c. The presence of distinct and adjustable resonance dips in each topological bandgap,
indicates the excitation of a resonance mode that is coupled to the edge state. In Fig. \ref{Fig2}d, a close-up view of the transmission and the theoretical fit for a selected mode is shown. This mode will be utilized for resonance-enhanced non-classical light generation.

\section{Enhancement the Nonclassical emissions using FDMR}
To generate entangled photon pairs using FDMR, we use a continuous-wave pump beam with a frequency that corresponds to one of the FDMRs with a $Q=5.2 \times 10^4$, as indicated by an arrow in Fig. \ref{Fig2}c. We note that the intrinsic quality factor of a single ring resonator for our devices $Q_i\approx 1.3\times10^4$ which is approximately one order of magnitude less than the measured intrinsic quality factor of FDMR $Q_i\approx 1.03\times10^5$. Maintaining energy conservation via the SFWM process requires the generation of photon pairs for idler and signal modes at different $\Delta_\mathrm{FDMR}$ symmetrically distributed around the pump frequency. To measure photon pairs that fulfill this condition and suppress the emission from FDMR modes in different $\Delta_\mathrm{FDMR}$, we utilize two narrowband tunable filters. These filters have center wavelengths placed at $6\Delta_\mathrm{FDMR}$, both below and above the pump wavelength, see Fig. \ref{Fig2}c. This setup enables efficient selection of the desired entangled photon pairs while minimizing interference from unwanted FDMR modes at different $\Delta_\mathrm{FDMR}$. We verify the non-classical nature of the photon pairs generated from FDMR by measuring the second-order cross-correlation function defined by $g_{si}^{(2)}(t)=\frac{\langle \hat n_s \hat n_i\rangle}{\langle\hat n_s \rangle. \langle \hat n_i \rangle }$, where $\hat n_{s/i}$ is the photon number of the signal/idler modes. This function determines the normalized probability of detecting signal and idler photons at a specific time separation $t$ and can be measured using coincidence rates of the signal-idler pairs and individual photon counts employing the SNSPDs and the time controller (TC). The second-order correlation function can be written in terms of measured total coincidence rate   of detecting signal and idler with a time separation of t, $N_{\text{tot},c}(t)$,
and measured total signal/idler count rates  $N_{\text{tot},s/i}$ (see Supplementary Information) \cite{Cruz2022}:
\begin{equation}
    g_{si}^{(2)}(t)=\frac{N_{\text{tot},c}(t)}{ N_{\text{tot},s} N_{\text{tot},i}}\times\frac{1}{T_{\mathrm{coin}}}
    \label{Eq: Second order correlation}
\end{equation}
where  $T_{\mathrm{coin}}$ is the duration of arrival time called coincidence window. We analyzed the second-order cross-correlation function $g_{si}^{(2)}(t)$, within the smallest measurable coincidence window, 100 ps, of our time controller.
\begin{table*}
  \centering 
  \caption{
Comparing the topological quantum photon-pairs sources 
achieved through SFWM in the SOI platform. The comparison is based on the CAR, pair generation rate (PGR), and brightness (B). The value of brightness is estimated using  $^{(a)}$the bandwidth of the edge state and $^{(b)}$the bandwidth of the coincidence counts (21pm). } 
\begin{tabular}{|l|c|c|c|c|}
\hline
\textbf{Quantum photon-pair sources based on TPIs} & \textbf{Power (mW)} & \textbf{CAR} & \textbf{PGR (MHz)}& \textbf{B (MHz/mW.nm)}  \\ 
\hline
Ref. \cite{Mittal2018} &1.4 &42 &  0.00015 & $<0.001^{(a)}$ \\
\hline
 \textbf{This work (FDMR)}& 0.086 & 2331& 0.031 & $17.165^{(b)}$ \\
 \hline
 \textbf{This work (FDMR)}&1.4 & 100 & 4.5 &  $153.061^{(b)}$\\
\hline
\end{tabular}
\label{table1}
\end{table*}

Figure \ref{Fig3}a shows the measured second-order cross-correlation function, with a maximum value of  $g_{si}^{(2)}(t) \approx 1450\pm50$ observed at $t=0$ and at a fixed pump power of $0.29$ mW, compelling evidence of time-correlated photon pairs at the FDMR.  At the same power, we compare these results with the second-order autocorrelation functions, $g_{ii}^{(2)}(t)\approx1$ and $g_{ss}^{(2)}(t)\approx1$, for the idler and signal modes, shown in Fig. \ref{Fig3}b. 
This comparative analysis allows us to examine a formal proof of a non-classical light source i.e. the violation of the Cauchy-Schwarz inequality, $\big[g_{si}^{(2)}(0)\big]^2 \leq \big[g_{ss}^{(2)}(0) \cdot g_{ii}^{(2)}(0)\big]$. Alternatively, we define and measure the nonclassicality parameter 
\begin{equation}
\Gamma:=\frac{\big[g_{si}^{(2)}(0)\big]^2} {\big[g_{ss}^{(2)}(0)\cdot g_{ii}^{(2)}(0)\big]}, 
\end{equation}
for which $\Gamma>1 $ indicates a nonclassical source. Figure \ref{Fig3}c illustrates measurements of the parameter $\Gamma$ as a function of the pump power and includes a comparison with the photons generated from the edge state when the FDMR mode is absent. The plot shows the violation of the Cauchy-Schwarz inequality ($\Gamma>1$), indicating a profound enhancement of nonclassical properties in the emitted photons due to the presence of the FDMR and its coupling to the edge state, which effectively increases the cross-correlation. At low powers, the experimental result is in good agreement with the theoretical model we developed in the Supplementary Information. However, at higher powers, the theory model deviates from the experimental data mainly due to the appearance of nonlinear effects, such as two-photon absorption (TPA) \cite{guo2018generation}, thermo-optic effects \cite{zimmerling2022broadband}, and thermal instability in the FDMR, which change the single and coincidence count rates. 
We note that the small size of our chip limits the edge state's ability to generate significant photon pairs while propagating at the shorter arm of the sample. Hence, unlike Ref. \cite{Mittal2018} the measured emission from the edge state is primarily composed of pump leakage and lacks considerable nonclassical properties i.e. $\Gamma\approx 1$.

To investigate the resonance-enhancement of the FDMR, at the pump power of 0.29 mW, the wavelength of the pump was swept across one resonance, allowing for the coincidence rate $N_c$, in the full-width half maximum (FWHM) of coincidence peak  $T_{\mathrm{coin}}=300$ ps (see Supplementary Information), to be compared to the transmission of the FDMR, as shown in Fig. \ref{Fig3}d. The peak of coincidence rate, pink trace, aligns with the deepest point of the resonance, purple trace, demonstrating successful resonance enhancement of the generation of non-classical photon pairs, with coincidence rate up to $\approx$ 400 kHz and
second-order cross-correlation up to $\approx$1450 (Fig.~\ref{Fig3}a). We measured the peak of the second-order cross-correlation $g_{si}^{(2)}(0)$ at different $\Delta_\mathrm{FDMR}$, as shown in Fig. \ref{Fig3}e. Different FSRs display different cross-correlation functions, mainly due to the variations in the total quality factors of the pump, idler, and signal modes. We note that this figure only shows the behavior of the cross-correlations at different FSRs and does not offer any information about frequency domain entanglement. Therefore, further measurements and characterization are needed in the future to clearly verify frequency entanglement in the current system.

Another important parameter for evaluating the nonclassical properties of the generated photon pairs is the Coincidence to Accidental Rate (CAR), defined by \cite{Caspani2017}
\begin{equation}
    \text{CAR}=\frac{\int_{-t/2}^{t/2}g^2_{si}(t)dt}{\int_{\infty-t/2}^{\infty+t/2}g^2_{si}(t)dt}.
\end{equation}
which serves as a reliable indicator of the signal-to-noise ratio of the non-classical photons and is obtained by integrating $g_{si}^{(2)}(t)$ over the peak at $t=0$. The CAR is measured as $\mathrm{CAR}\propto \frac{N_{tot,c}}{N_{tot,s} N_{tot, i}}$,
Note that, all three count rates  $N_{tot,j}$ 
 ($j=c, s, i$) scale quadratically with the pump power (see Supplementary Information), resulting in the observed inverse dependence of CAR with the square of power in agreement with our theoretical model, see Fig. \ref{Fig3}f. This figure also provides a comparison of the CAR between the FDMR and the edge state (heater off) under various pump powers. Our source achieves an exceptional CAR value of $2300$, significantly surpassing the $\mathrm{CAR}\approx 1.01$ obtained with edge states in the absence of the FDMR. 

\begin{figure*}[t]
    \centering
 \includegraphics [width=0.982\linewidth]{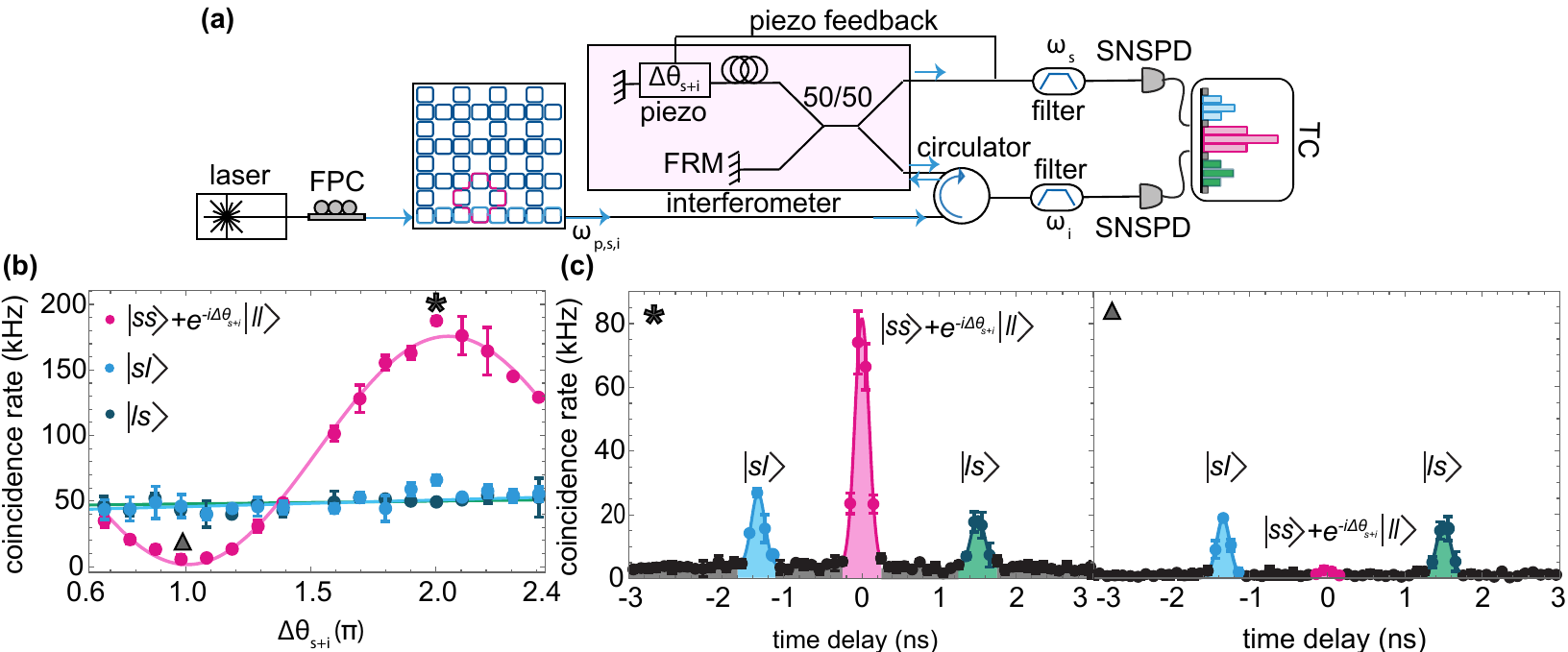}
    \caption{\textbf{Time-energy entanglement.}   \textbf{a}, Measurement setup for a folded Franson interferometer, including a 50:50 beamsplitter, Faraday rotator mirror (FRM),  circulators to separate the input and output of the interferometer, and a stabilization system. \textbf{b}, Coincidence rates versus the relative phase difference \(\Delta\theta_{s+i}\) for indistinguishable states \((|ss\rangle + e^{i\Delta\theta_{s+i}}|ll\rangle)\) and distinguishable states \(|sl\rangle\) and \(|ls\rangle\), represented by pink, blue, and dark green points, respectively. The duration of arrival time
    is $T_{coin}=400$ ps. The data is fitted with the function \(\alpha + \beta \cos(\Delta\theta_{s+i})\) (solid lines), where \(\frac{\beta}{\alpha}\) gives the two-photon interference visibility of $0.982$, and exceeds 0.99 after noise subtraction.  \textbf{c}, Coincidence rates versus arrival time for the two points marked in plot \textbf{b} by a star and a triangle, indicating the highest and lowest coincidence rates of indistinguishable states. These coincidence rates are obtained from measured coincidence counts during an acquisition time of 5 seconds.}
    \label{Fig4}
\end{figure*}

The result highlights the outstanding performance of the FDMR in enhancing the nonclassical characteristics of the generated photons. In Table \ref{table1}, we compare our resonance-enhanced Floquet Topological quantum source and the other
topological quantum source using edge state in silicon microrings \cite{Mittal2018}. Compared with basic microring resonators \cite{ Kumar:13, guo2018generation, clementi2023programmable}, FDMR can be turned on and off in situ while also having the capability to couple with edge states. Multiple FDMRs can be activated and coupled everywhere in the lattice located on a single chip.  These characteristics distinguish our system from the traditional approach of entanglement generation using microring resonators.

\section{Time-energy Entanglement}
Our topological sample can be considered as a narrowband time-energy entangled source, making it potentially suitable for long-distance quantum communication purposes 
 \cite{PhysRevLett.93.180502, Yu2020}. In this section, we demonstrate that the photon pairs generated from FDMR are indeed entangled. We verify this entanglement by measuring the visibility of two-photon interference using a folded Franson interferometer \cite{PhysRevLett.62.2205, PhysRevLett.65.321, steiner2021ultrabright, rahmouni2024entangled}. To implement the folded Franson interferometer, we utilize a fiber-based unbalanced Michelson interferometer with a path delay of $\tau_d = 1.44$ ns, as shown in Fig.~\ref{Fig4}.a and explained in the Supplementary Information. This path delay is selected to be less than the coherence time of the pump laser, which has a $200$ kHz bandwidth, and greater than the coherence time of the bi-photon ($\tau_c \approx 60$ ps, derived from the bandwidth of the generated photon pairs), thus avoiding first-order interference \cite{PhysRevLett.62.2205}.  
 
Each photon can pass the interferometer via either the short $(s)$ or long $(l)$ paths, resulting in four possible outcomes for each signal-idler pair at the output. To measure the visibility of the two-photon interference, we send the generated signal-idler pairs to one input of the interferometer. When the signal and idler photons travel through different arms, the states $|sl\rangle$ and $|ls\rangle$ are created and detected by SNSPDs, with an arrival time delay of $\pm \tau_d$ corresponding to the path-length difference. These possibilities produce two side peaks in the coincidence histogram, highlighted by the blue and green sections in Fig.~\ref{Fig4}c.
If both signal and idler photons pass through the short or long arm of the interferometer, an entangled state is generated as $(|ss\rangle + e^{i\Delta\theta_{s+i}}|ll\rangle) / \sqrt{2}$,
where $\Delta\theta_{s+i}$ is the phase acquired by signal and idler when they propagate over the long arm.

The two-photon interference (and thus time-energy entangled) can be observed (verified) by adjusting the relative phase of the interferometer, as shown in Fig. \ref{Fig4}b. The pink points are fitted by the function $\alpha + \beta\, \text{cos}(\Delta\theta_{s+i})$, where \(\frac{\beta}{\alpha}\) indicates the two-photon interference visibility of $0.982$, which exceeds 0.99 after noise cancellation. This visibility is well above the separability threshold of $0.707$, thereby violating the Clauser–Horne–Shimony–Holt inequality \cite{Clauser}. Figure \ref{Fig4}b also shows the coincidence rates for the separable states $|sl\rangle$ and $|ls\rangle$, represented by blue and dark green points, respectively, which remain nearly constant as the phase varies. Figure \ref{Fig4}c illustrates the coincidence rate versus the arrival time of photons corresponding to the two points marked by a star and a triangle in Fig. \ref{Fig4}b. The values of the coincidence rates in Fig. \ref{Fig4} have been adjusted to account for the loss from the silicon waveguides inside the chip to the SNSPDs.

\section{Resiliency against defect}

We emphasize that the nature of FDMR mode comes from the topologically nontrivial behavior of the lattice, in that, the existence of circulating loops in the bulk of the lattice \cite{Afzal2021, mukherjee2020observation} is required to mimic the quantum hall effect that causes light to propagate only at the edge of the lattice unidirectionally. In fact, the topology of the lattice makes FDMR loops robust to fabrication imperfections as long as the defect doesn't completely alter the hopping sequence, for instance, by
removing one ring in the loop where the FDMR is excited.

\begin{figure}[t]
    \centering
 \includegraphics [width=0.9\linewidth]{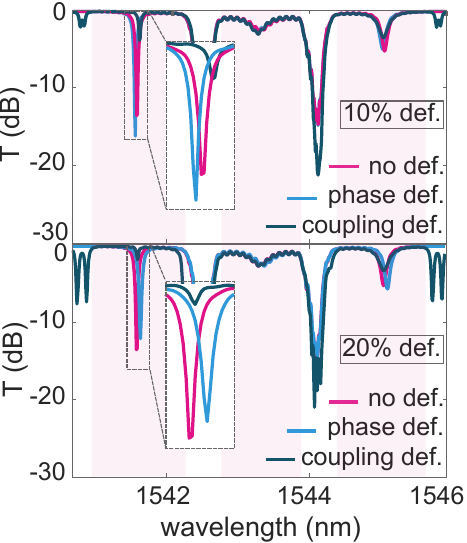}
    \caption{\textbf{Robustness of FDMR vs fabrication imperfections.} \textbf{a} and \textbf{b,} Simulation of respectively $10\%$ and $20\%$  variations in coupling coefficient between microrings and roundtrip phase of the microrings located around the FDMR resonance. Pink lines represent the lattice without defects. Light blue and dark green lines show the maximum changes in FDMR due to coupling defects and roundtrip phase defects, respectively. The maximum changes, in terms of both frequency shift and resonance linewidth, are chosen from 50 simulations with random defects in the coupling coefficient representing the variations in the gaps between microrings and roundtrip phase defects which are due to variations in the size of the microrings. As illustrated in this figure the defects in the roundtrip phases cause shifts in resonance and defects in coupling coefficient $p_c$ lead to changes in the Q-factor of the FDMR. }   
    \label{Fig2methods}
\end{figure}

To illustrate the resilience of FDMRs, we conducted simulations of the lattice's transmission spectrum, considering deviations of $10\%$ and $20\%$ in coupling coefficients between microrings and variations in the roundtrip phase of each microring. Our simulations encompassed 50 lattices, incorporating random variations centered around the FDMR loop. The most significant impacts on the transmission spectrum are presented in Fig.~\ref{Fig2methods}, shown by the dark green and light blue lines for coupling coefficient and roundtrip phase variations, respectively. This illustration confirms that specific types of defects, whether within the FDMR loop or adjacent microrings, manifest as loss channels. While these defects may alter the quality factor of the FDMR resonance or cause a resonance shift, they are unable to destroy the FDMR loop.

As discussed above, various forms of defects and disorders can appear within our devices. This section aims to elaborate further on distinct scenarios that could potentially impact the efficiency of our system.

\textbf{1. Defect affecting the edge state:} A possible scenario involves the emergence of defects/disorders along the trajectory of the edge state, far away from the region where the FDMR is formed. In such an instance, the system demonstrates resilience in the face of defects. The spectral characteristics of the system remain largely unaffected by the presence of such a defect \cite{Dai2022}. It is worth noting that this resilience pertains specifically to certain types of defects, as the edge state remains robust against select anomalies, as long as the defects/disorders do not cause the band gap to close. Examples of such defects are variations in the microring round trip phases, couplings, and scattering loss due to sidewall roughness in the waveguides.

\textbf{2. Impact on neighboring unit cells of the FDMR:} In this scenario, the defects surround the FDMR mode, yet they do not directly intersect the FDMR loop itself. Instead, the adjacent unit cells function as loss channels, taking energy and photons away from the FDMR mode. Consequently, the introduction of these neighboring defects results in increased intrinsic losses for the mode. Such imperfections in fabrication can manifest in various ways, such as by fluctuations in the spacing between rings composing the FDMR and those situated within neighboring unit cells. These variations exert a direct influence on the overall quality factor of the mode. It's important to highlight that the mode remains resilient against these defects, provided they do not exceed a certain magnitude that causes the band gap to close. This assertion finds support in our simulation results, depicted in Fig. \ref{Fig2methods}.

\textbf{3. Defects impacting the FDMR loop:} 
The third category encompasses defects that directly affecting on the rings forming the FDMR. This type of defect has the most significant impact on the FDMR. It shares similarities with the second type, potentially inducing frequency shifts due to disordered configurations and resulting in phase alterations. Moreover, these defects have the capacity to influence the quality factor of the FDMR mode by introducing loss channels through light scattering, such as caused by surface roughness or coupling variations. However, in the case of a significant deviation in the Q-factor of the FDMR, one can turn off the affected FDMR and excite another FDMR far from the defect, as FDMRs can occur anywhere in the lattice and can be easily turned off and on. While for other topologically trivial resonators, such as single-ring resonators, there is no possibility to deactivate the affected resonator and activate another resonance.

Another consequence involves the potential disruption of extrinsic coupling with the edge state. Such disruptions can yield variations in extrinsic quality factors, either increasing or reducing them. These defects, although impactful, do not have the capability to destroy the entire mode, as long as the defect's size remains proportionally smaller than that of a single ring within the loop. Figure \ref{Fig2methods} shows simulations depicting these specific defects. Furthermore, we note that the mode's inherent frequency tunability plays a crucial role in mitigating the effects of defects, particularly those resulting in phase shifts within the loop. We finally note that in topologically trivial structures, such as a single ring resonator coupled to a waveguide, a defect or obstacle in the waveguide or resonator can block the propagation path or completely destroy the resonance mode. This can significantly affect photon generation rates, CAR, and entanglement. While with our topological system, both the edge state and FDMR mode can bypass the defect and continue propagation along an alternative path, maintaining the resonance mode. Consequently, our device can be considered as a robust nonclassical source of photon pairs.

\section{Summary and Discussions}
In summary, we have demonstrated the resonance-enhanced generation of entangled photon pairs in anomalous Floquet TPI, operating at room temperature by employing an FDMR coupled to a topological edge mode.
To verify the nonclassicality of the generated entangled photons, we measured the second-order cross-correlation, tested the nonclassical characteristic of the emission, and provided a comparison with the comprehensive theoretical model that we have developed. Using the folded Franson interferometer, we also verified the time-energy entanglement of the generated photon pairs, achieving a pair photon interference visibility of $98.2\%$.

We conclude our discussion by exploring the topological protected properties of our system.  Unlike previous approaches that depend on propagating fields in the edge state to generate photon pairs, our approach distinctively employs a confined mode embedded within the bandgap region co-existing with the edge state. Therefore, the FDMR operates as a frequency-tunable source that generates entangled pairs, while the edge state acts as a topologically protected waveguide that delivers the generated photons to the output ports. Consequently, our system acquires the same topologically protected features as those inherently present within the edge state. This unique capability distinguishes the current system from other sources of entanglement that rely on edge states \cite{Mittal2018, Mittal2021HOM, Dai2022}, and potentially from conventional photonic devices as well \cite{Kumar:13, ma2017silicon,ma2020ultrabright, afifi2021contra, guo2018generation, clementi2023programmable}.
We note that defects and disorders at the physical boundaries, where the FDMR forms, have the potential to interfere with the loop,
shifting the resonance mode.
However, the frequency tunability of the FDMR can partially mitigate these effects. 

An essential aspect of our sample is its capability to generate entanglement within a relatively small chip due to the compact size of the FDMR. This feature becomes especially crucial when the material loss of the sample affects the nonclassical property of the photon pairs. The compact nature of the FDMR allows for efficient entanglement generation and on-chip distribution, making it well-suited for scenarios where larger devices might encounter challenges related to propagation loss. The unique properties of the FDMR, including its high quality factor, compact size, tunability, and ability to couple with edge states, play a critical role in facilitating the generation of bright entangled photon pairs. Using FMRs can open up possibilities for exploring novel techniques of on-chip entanglement distribution, field-matter interaction, and advancing integrated quantum electrodynamics. In particular, the FDMR can efficiently interact with localized atoms/ions in the substrate, facilitating their integration into a network of cascaded atom-cavity chains connected with edge states. This platform holds promise for the development of on-chip and topologically-protected quantum computing and processing.

\textbf{Acknowledgments} We thank Mohammad Hafezi, Rogério de Sousa, and Daniel Oblak for helpful comments and discussions, and Tyler Zegray for preparing the figures. S.B. acknowledges funding by the Natural Sciences and Engineering Research Council of Canada (NSERC) through its Discovery Grant, funding and advisory support provided by Alberta Innovates (AI) through the Accelerating Innovations into CarE (AICE) -- Concepts Program, support from Alberta Innovates and NSERC through Advance Grant project, and Alliance Quantum Consortium. V.V. and T.J. acknowledge funding from NSERC and AI.  This project is funded [in part] by the Government of Canada. Ce projet est financé [en partie] par le gouvernement du Canada. 
\\




\onecolumngrid
\appendix
\begin{center}
\newpage
\textbf{\large Supplementary Materials}
\end{center}
\subsection{2D  microring lattices as periodically-driven (Floquet) systems}

The Floquet-Bloch Hamiltonian of our 2D microring lattice can be derived using a similar approach as for the general 2D microring lattice in \cite{Afzal2018}, except here each unit cell contains three microrings instead of four. We first transform the microring lattice into an equivalent 2D coupled waveguide array by "cutting" each microring in a unit cell and unrolling it into a straight waveguide, as shown in Fig.~\ref{Mapping}, with length $L$ equal to the circumference of the microring. Thus, each roundtrip of light circulation in a microring is equivalent to light propagation along the coupled waveguide array of length $L$, which can be divided into a sequence of 4 coupling steps of equal length $L/4$. Denoting the fields in microrings A, B, and C in unit cell $(m,n)$ in the lattice as $\psi_{m,n}^A, \psi_{m,n}^B$ and $\psi_{m,n}^C$, respectively, we can write the equations for the evolution of light in the waveguide array along the direction of propagation ($z$-axis) over each period in terms of the coupled mode equations:
\begin{figure*}[ht]
  \centering
  \includegraphics[width=1\textwidth]{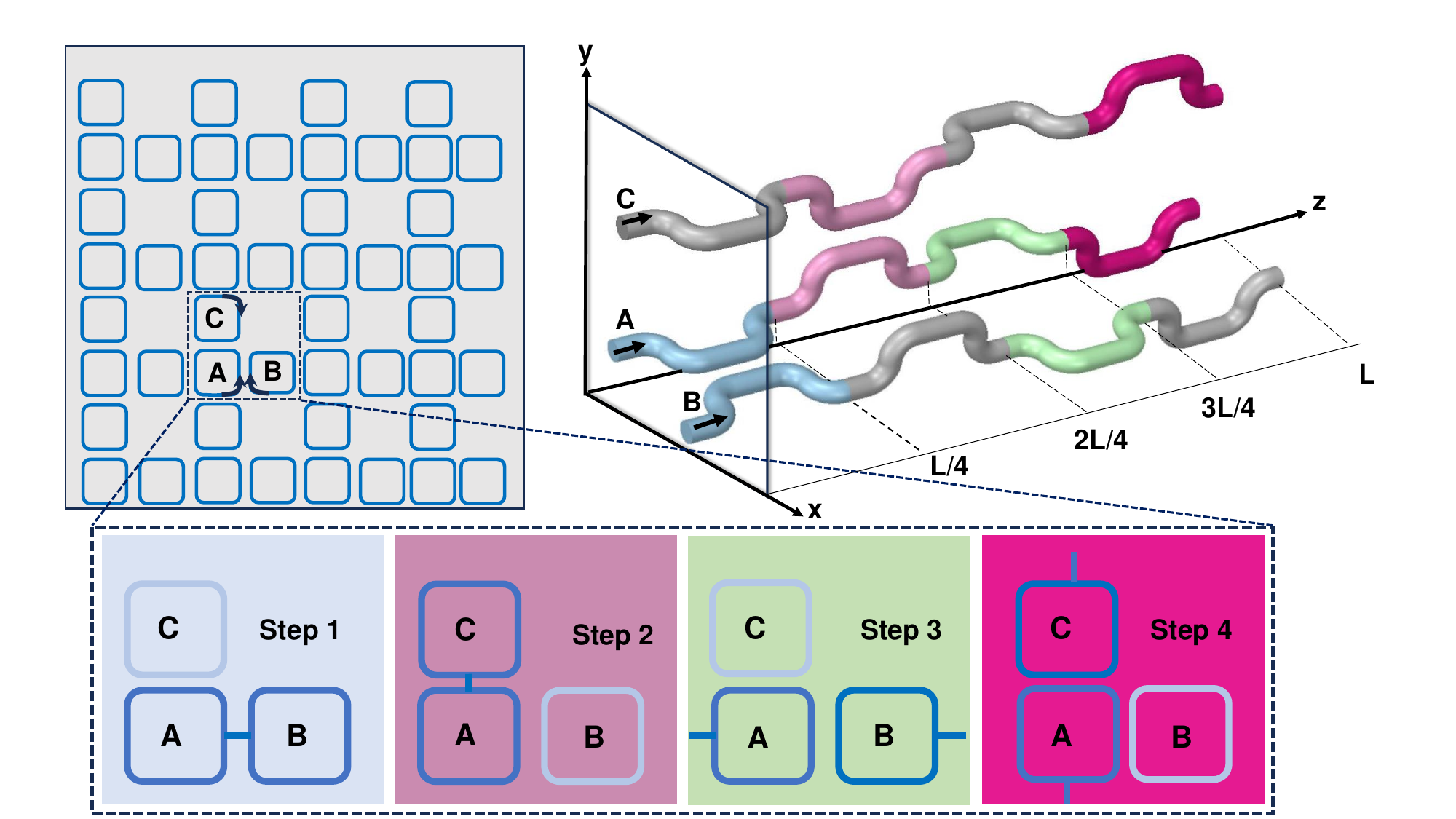}
  \caption{Mapping the coupling sequence of a unit cell of 2D microring lattice into a coupled waveguide array in the direction of propagation ($z$). The coupled waveguide array is obtained by unrolling the microrings in a unit cell from the points shown by black arrows. }
\label{Mapping}
\end{figure*}

\begin{eqnarray*}
 -i\frac{\partial{\psi_{m,n}^A}}{\partial{z}} = \beta\psi_{m,n}^A + 
    k(1)\psi_{m,n}^B + k(2)\psi_{m,n}^C 
    + k(3)\psi_{m-1,n}^B + k(4)\psi_{m,n-1}^C
\end{eqnarray*}
\begin{eqnarray*}
   -i\frac{\partial{\psi_{m,n}^B}}{\partial{z}} = \beta\psi_{m,n}^B + k(1)\psi_{m,n}^A + k(3)\psi_{m+1,n}^A 
  \end{eqnarray*}
  \begin{eqnarray}
  -i\frac{\partial{\psi_{m,n}^C}}{\partial{z}} = \beta\psi_{m,n}^C + k(2)\psi_{m,n}^A + k(4)\psi_{m,n+1}^A 
   \label{eq_motion}
  \end{eqnarray}
where $\beta$ is the propagation constant of the microring waveguide mode and $k(j)$ specifies the coupling strength per unit length between adjacent waveguides in step $j$ of each evolution period, with $k(j)=\theta/(L/4)=\arcsin(p_c)/(L/4)$ in step $j$ and $0$ otherwise ($\theta$ and ${p_c}^2$ are respectively the coupling angle and the percentage of the power coupled between adjacent microrings). By making use of the Bloch's boundary conditions $\psi_{m+1,n} = \psi_{m,n}e^{ik_x\Lambda}$, $\psi_{m,n+1} = \psi_{m,n}e^{ik_y\Lambda}$, where $\Lambda=2a$ is the lattice constant  ($a$ is the distance between the centers of adjacent microrings), we can express the coupled mode equations in the form of
\begin{equation}
-i\frac{\partial}{\partial z} |\psi(\textbf{k}, z)\rangle = [\beta I + H_{FB}(\textbf{k},z)]|\psi(\textbf{k}, z)\rangle 
\label{eq:S4}
\end{equation}
where $|\psi\rangle = [\psi_{m,n}^A,\psi_{m,n}^B,\psi_{m,n}^C]^T$, $\textbf{k}=(k_x,k_y)$ is the crystal momentum vector, and $I$ is the identity matrix. The Floquet-Bloch Hamiltonian $H_{FB}$ consists of a sequence of 4 Hamiltonians $H_{FB}= \sum_{j} (\Theta(z-(j-1)L/4)-\Theta(z-jL/4) ) H(j)$, with $\Theta(z)$ the step function, corresponding to coupling steps $j = \{1,2,3,4\}$,  which have the explicit forms for $j = {1, 3}$ 
\begin{eqnarray*}
H(j) =& 
 \Bigg(
    \begin{matrix*}[c]
    0 & k e^{-i \delta_{3j} k_x \Lambda} & 0\\
    k e^{i \delta_{3j} k_x \Lambda} & 0 & 0\\
    0 & 0 & 0
    \end{matrix*}
    \ \Bigg)
\end{eqnarray*}
and for $j = {2, 4}$
\begin{eqnarray*}
H(j) =&
 \Bigg(\ 
    \begin{matrix*}[c]
    0 & 0 & k e^{-i \delta_{4j} k_y \Lambda}\\
    0 & 0 & 0\\
    k e^{i \delta_{4j} k_y \Lambda} & 0 & 0\\
    \end{matrix*}
    \Bigg) \\
\end{eqnarray*}
In the above expressions, $\delta_{ij} = 1$ if $i = j$ and 0 otherwise.  The Floquet-Bloch Hamiltonian is periodic in $z$ with a periodicity equal to the microring circumference $L$, $H_{FB}(\textbf{k},z)=H_{FB}(\textbf{k},z+L)$. The quasienergy spectrum of the lattice consists of three bands in each Floquet-Brillouin zone, with a flat band state at quasienergies $2m\pi,\  (m\in \mathbb{Z})$. As shown in Ref.~\cite{Afzal2018}, for weak coupling strengths,  $\theta < \pi/\sqrt{8}$, the microring lattice has energy bands with nontrivial Chern numbers as well as one conventional insulator bandgap (with trivial winding number) and two symmetric bandgaps with nontrivial winding numbers, called Floquet Chern insulators.
For the strong coupling regime ($\theta \gtrsim \pi/\sqrt{8}$), all the energy bands have trivial Chern numbers, while the three bandgaps exhibit anomalous Floquet insulator behavior with nontrivial winding numbers.
\par
\subsection{Theory and topological features of Floquet Defect Mode Resonance (FDMR)}
The FDMR is created by introducing a periodic perturbation to the Hamiltonian of the topological microring lattice in the form of a phase detune of a microring during the step $j$ in each evolution period.  Thus, if a phase detune of $\Delta \phi$ is applied to microring B in a unit cell during step $j$, its equation of motion is modified as \cite{Afzal2021}
\begin{equation}
    -i\frac{\partial{\psi_{m,n}^B}}{\partial{z}} = (\beta+ \Delta \beta(j))\psi_{m,n}^B + k(1)\psi_{m,n}^A + k(3)\psi_{m+1,n}^A 
    \label{eq:S5}
\end{equation}
where $\Delta \beta(j) = 4\Delta \phi /L$ in step $j$ and zero otherwise. We previously showed in Ref. \cite{Afzal2021} that unlike static defect modes created in conventional bandgaps, the spatial field localization pattern of the defect mode depends on which segments of the microring are detuned. For instance, if the detune is applied to a microring A located in the bulk of the topological lattice, two or four coupled FDMR loops can get excited depending on which microring segment the detune is applied to. To create only one FDMR loop, we need to apply the phase detune to segment $j = 2 (1)$ or $4 (3)$ of microring B (C), otherwise two coupled FDMR loops located at the top (right) and bottom (left) of microring B (C) will be excited. In our experiment, to couple the FDMR to an edge mode along the bottom boundary, we apply the detune to a microring B located on the bottom boundary.  The presence of the boundary also ensures that only one FDMR loop (instead of two in the bulk) will be excited.  Thus, for simplicity, we can detune the whole microring B to excite the single FDMR. It is worth mentioning that detuning the whole microring A at the boundary of the lattice will excite two coupled FDMRs. Hence, for the purpose of exciting a single FDMR, it becomes necessary to selectively detune solely either segment $j=1$ or $j=3$ of microring A at the bottom boundary of the lattice.  However, implementing this step experimentally poses a challenge due to the intricate requirement of a minute heating apparatus, situated atop a designated one-quarter section of microring A.

\par
\begin{figure*}[ht]
  \centering
  \includegraphics[width=0.55\textwidth]{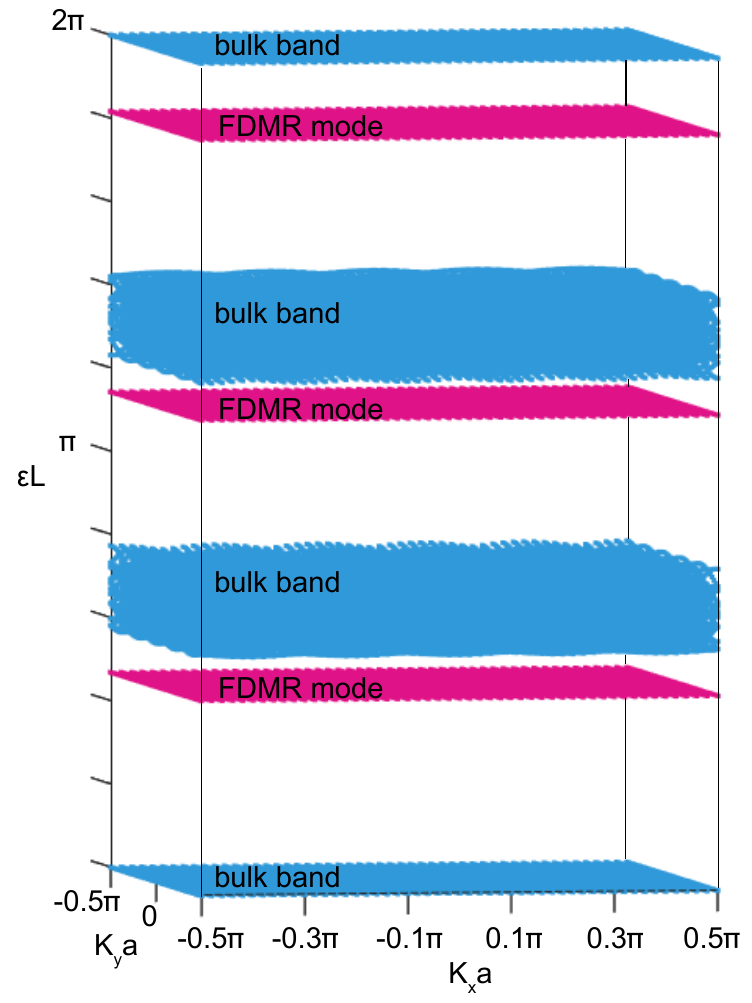}
  \caption{The band structure of a supercell, wherein a phase detuning $\Delta \phi = 0.55\pi$ has been selectively imposed upon the sequence $j=4$ of microring B located at the center of a $5\times5$ unit cells.  The periodic boundary condition in both $x$ and $y$ directions has been adapted to compute this 3D band diagram.}
\label{Fig:Supercell}
\end{figure*}

We can first construct the Hamiltonian matrix $H_j(k_x)$ for each coupling step $j = {1, 2, 3, 4}$ for the $1 \times N_y$ lattice with periodic boundary condition on $x$ and zero open boundary condition on $y$ (detailed matrix construction is given in \cite{Afzal2018}). The quasienergy bands over one Floquet-Brillouin zone can be obtained by computing eigenvalues of the Floquet operator $U_F(k_x) = U_4(k_x)U_3(k_x)U_2(k_x)U_1(k_x)$, where $U_j(k_x) = e^{-i H_j(k_x) L/4}$, shown in Fig. 1c of the main manuscript. To compute the quasienergy band for the FDMR at a specific phase detune $\Delta\phi$, we can construct the Hamiltonian matrices $H_j(k_x,k_y)$ for a supercell consisting of $(N_x = 5\times N_y = 5)$ unit cells with periodic boundary conditions on all boundaries (detailed matrix construction is given in the Supplementary Material Section S1 of \cite{Afzal2021}). With the phase detune applied to coupling step $j=4$ of ring B located at the center of the supercell, we can obtain the quasienergy bands of the FDMR modes over one Floquet-Brillouin zone by computing the eigenvalues of the Floquet operator $U_F$ of the supercell. Figure~\ref{Fig:Supercell} illustrates the obtained energy bands of the supercell, where the blue bands denote transmission bands, and the flat pink bands signify the eigenenergies associated with FDMRs. Leveraging the outcomes obtained from the supercell analysis, we proceeded to overlay the eigenenergies of the FDMRs onto the band diagram of a semi-infinite microring lattice. This representation is depicted in Fig. 1c of the main manuscript.

\par
To obtain the intensity distributions of the edge mode and FDMR, shown in Fig. 2b of the main manuscript, we simulated a $(N_x = 5\times N_y = 5)$ lattice with open boundary conditions on all boundaries. With an input waveguide coupled to the bottom left ring A and an output waveguide coupled to the bottom right ring B, a continuous-wave light signal applied to the upper port of the input waveguide allows for the computation of the field amplitudes in all the microrings in the lattice using the field coupling method presented in \cite{Tsay2011}.

\subsection{Topological FDMR and nontopological defect mode resonance}
\begin{figure*}[ht]
  \centering
  \includegraphics[width=1\textwidth]{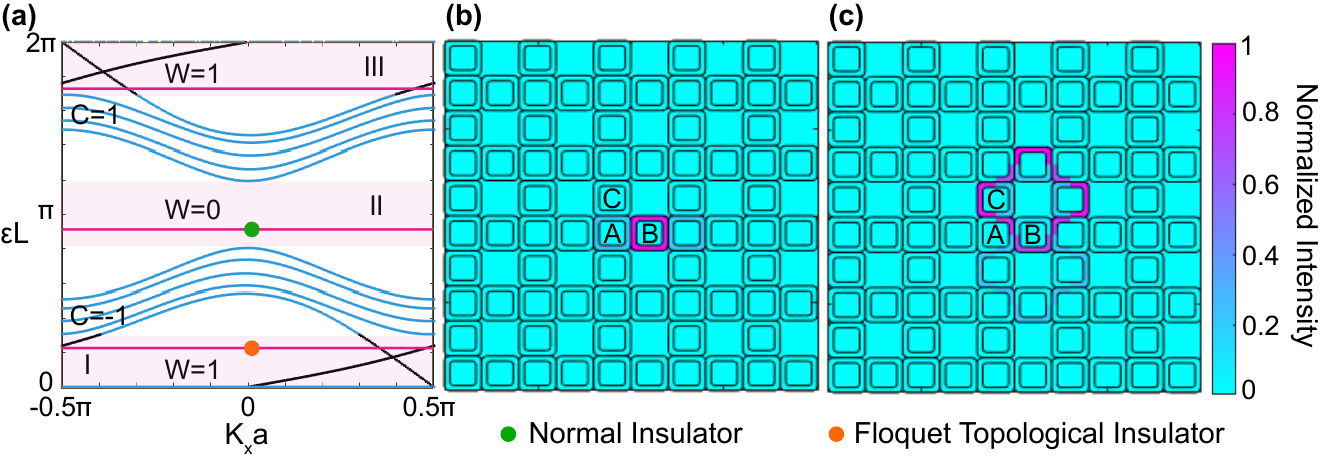}
  \caption{\textbf{a}, Band diagram of a semi-infinite microring lattice characterized with coupling power of $65\%$. Highlights illustrate three bandgaps while black lines indicate edge states in topologically nontrivial bandgaps.  Pink lines, mapped from a supercell with applied phase detune $\Delta\phi=1.2\pi$ to the sequence of $j=4$ of microring B, show the defect modes. \textbf{b and c,} Represent the field distribution of the normal defect mode resonance and FDMR, respectively, in conventional bandgap and topological bandgap.}
\label{Fig: Comparing_FDMR_with_Normal_Resonance}
\end{figure*}

As previously shown in Ref.~\cite{Afzal2021}, the emergence of FDMR is intrinsically tied to the topological nature of the bandgap and can not be created in the conventional bandgap. To have a better understanding of the topological essence inherent to FDMRs, we theoretically showed the difference between the localization pattern of FDMR and a conventional defect mode resonance. In this simulation, we consider a microring lattice characterized with coupling angle $\theta=0.3 \pi$ (corresponding to a coupling power of ${p_c}^2=65\%$). Within this context, a notable feature emerges—both topologically nontrivial and conventional (trivial) bandgaps manifest within a singular Floquet-Brillouin zone. This phenomenon is depicted in Fig.~\ref{Fig: Comparing_FDMR_with_Normal_Resonance}a, where the eigenenergies of a semi-infinite microring lattice, consisting of $1\times5$ unit cells which are limited in $y$ direction and infinite in $x$ direction, are computed. The existence of clockwise and counterclockwise edge states (black lines) affirms the topological character of bandgaps I and III. To project the eigenenergies of defect modes onto the band diagram of a semi-infinite lattice, we simulated a supercell with periodic boundary conditions along $x$ and $y$ directions. The supercell, comprised of $5\times5$ unit cells, was subjected to a phase-detuning of $\Delta\phi=1.2\pi$ applied to the sequence $j=4$ of microring B at its central region. The resultant eigenenergies of the defect modes are depicted as pink lines in Fig.~\ref{Fig: Comparing_FDMR_with_Normal_Resonance}a. 

For the purpose of comparing the resonance patterns of defect modes within topological and conventional bandgaps, we plotted the field distributions of two eigenenergies indicated by green and orange circles on the band diagram. This assessment utilized eigenvalues derived from the supercell, where $K_x=K_y=0$. The outcomes of our simulations conclusively demonstrate that FDMRs, identifiable by their specific loop patterns wherein light resonates within a cavityless loop unrestricted by defect points, exclusively manifest within topologically nontrivial bandgaps (as illustrated in Fig.~\ref{Fig: Comparing_FDMR_with_Normal_Resonance}c. In stark contrast, the field distribution associated with the defect mode situated within a conventional bandgap reveals the confinement of the mode at the microring defect (also shown in Fig.~\ref{Fig: Comparing_FDMR_with_Normal_Resonance}b).

An important caveat lies in the fact that defect modes lacking topological attributes within a conventional bandgap remain experimentally inaccessible due to the absence of traveling modes conducive to their excitation and coupling. In contrast, the FDMR can be feasibly experimentally excited by harnessing an edge state that coexists within the same topological bandgap.

\subsection{Theoretical modeling of the pair generation}
\begin{figure*}[t]
  \centering
  \includegraphics[width=1\textwidth]{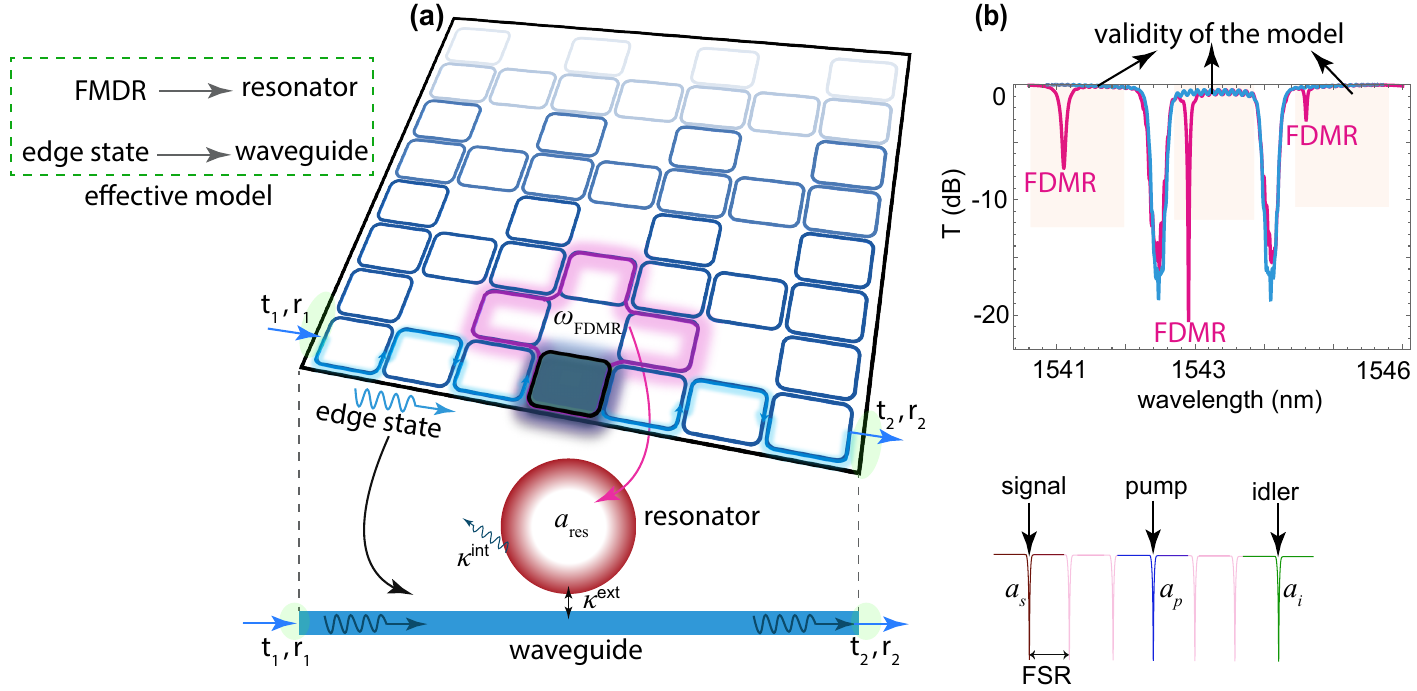}
  \caption{\textbf{a} The coupling between the FDMR and edge state can be represented as a single ring resonator with a radius of R coupled to a trivial waveguide. In this representation, the FDMR functions as a single-mode resonator with a resonance frequency $\omega_{\text{FDMR}}$, an annihilation operator $a_{\text{res}}$, and a Free-Spectral-Range (FSR). To initiate SFWM in this configuration, the system is pumped at mode $a_p$, resulting in the generation of photon pairs at the idler $a_i$ and signal $a_s$ modes, distributed above and below the pump mode. The input and output light undergo reflection and transmission at the facets of the waveguide with parameters $t_i$ and $r_i$. \textbf{b} The output of the FDMR indicates the regions where the effective resonator-waveguide model remains valid.}
\label{FigSI2}
\end{figure*}

Considering the FDMR as a local resonator with a resonance frequency $\omega_\mathrm{FDMR}$ and the edge state as a waveguide connected to the resonator with a coupling rate $\kappa_e$, we can simplify the representation of the FDMR-edge state system as a resonator-waveguide platform, as shown in Fig. \ref{FigSI2}a. This simplification provides a straightforward approach to effectively characterize and model the system's dynamics and describe the generation of nonclassical radiation from the FDMR. Note that, the validity of this model is confined to the 
topological bandgap, as illustrated in Fig. \ref{FigSI2}b. The phenomenological Hamiltonian governing the SFWM of the system in the effective representation is
\begin{equation}
    H=\sum_{j=p,s,i}\hbar\omega_j a_j^\dagger a_j-\hbar\big(g_{\text{nl}} a_p^2 a_s^\dagger a_i^\dagger+h.c), \label{Ham1}
\end{equation}
where $a_j$ with $j=p,s,i$ are the annihilation operators of pump ($p$), signal ($s$), and idler ($i$). The first term in the Hamiltonian (\ref{Ham1}) corresponds to the free energy of each mode, while the last term describes the Hamiltonian of SFWM with a nonlinear coupling parameter $g_{\text{nl}}=3 \hbar \omega_p^2 \chi^{(3)}/4\epsilon_0 \Bar{n}^4 V_{\text{FDMR}}$. In this expression, $\chi^{(3)}$ denotes the nonlinear susceptibility, $V_{\text{FDMR}}$ represents the mode volume of the FDMR, $\omega_p$ is the pump frequency, $\Bar{n}$ stands for the average index of refraction of the FDMR, and $\epsilon_0$ corresponds to the permittivity of free space.

The dynamics of the system can be described by the quantum Langevin equations of motion \cite{Cui2022, PhysRevLett.109.130503, Arnold2020}
\begin{eqnarray}\label{EQM1}
    \Dot{a_p}&=& -(i\omega_p+\frac{\kappa_p}{2})a_p+2ig_{\text{nl}}a_p^\dagger a_s a_i+\sqrt{\kappa_p^{\text{ext}}}a_p^{\text{ext}}+\sqrt{\kappa_p^{\text{int}}}a_p^{\text{int}}, \nonumber \\
     \Dot{a_s}&=& -(i\omega_s+\frac{\kappa_s}{2})a_s+ig_{\text{nl}}a_p^2 a_i^\dagger+\sqrt{\kappa_s^{\text{ext}}}a_s^{\text{ext}}+\sqrt{\kappa_s^{\text{int}}}a_s^{\text{int}},\\
     \Dot{a_i}&=& -(i\omega_i+\frac{\kappa_i}{2})a_i+ig_{\text{nl}}a_p^2 a_s^\dagger +\sqrt{\kappa_i^{\text{ext}}}a_i^{\text{ext}}+\sqrt{\kappa_i^{\text{int}}}a_i^{\text{int}}, \nonumber
\end{eqnarray}
where the total damping rate $\kappa_j=\kappa_j^{\text{ext}}+\kappa_j^{\text{int}}$ of each mode $(j=p, s, i)$ is the sum of intrinsic $\kappa_j^{\text{int}}$ and extrinsic $\kappa_j^{\text{ext}}$ damping rates. Furthermore, the term $a_j^{\text{ext}} (a_j^{\text{int}})$ accounts for the noise operator arising from the coupling to the edge state (inaccessible loss channels). When a strong laser pumps the system at the resonance frequency $\omega_p$, we can simplify the first equation in (\ref{EQM1}) by considering only the steady-state part of the pump and disregarding field fluctuations. In this scenario, the pump can be treated as a classical field with the amplitude
\begin{equation}
   \alpha_p (\omega)=| \langle a_p\rangle|\approx \sqrt{\frac{\kappa_p^{\text{ext}}}{(\omega-\omega_p)^2+\kappa_p^2/4}}\alpha_{in},
\end{equation}
where $\alpha_{in}=\sqrt{\frac{P}{\hbar \omega_{p}}}$ and $P$ is the power of the laser in the edge state. Note that, in the derivation of the above equation, we neglected the nonlinear term $2ig_{\text{nl}}\langle a_p^\dagger a_s a_i\rangle$ under the assumption of the non-depletion regime \cite{4wmixing_sipe}. Considering the pump as a classical field, we can solve the equations of motion for the idler and signal in the frequency domain. By taking the Fourier transformation of the second and third equations in (\ref{EQM1}), we obtain

\begin{eqnarray}
    -i \omega a_s&=& -(i\omega_s+\frac{\kappa_s}{2})a_s+iGa_i^\dagger+\sqrt{\kappa_s^{\text{ext}}}a_s^{\text{ext}}+\sqrt{\kappa_s^{\text{int}}}a_s^{\text{int}},\\
     -i \omega a_i&=& -(i\omega_i+\frac{\kappa_i}{2})a_i+iG a_s^\dagger +\sqrt{\kappa_i^{\text{ext}}}a_i^{\text{ext}}+\sqrt{\kappa_i^{\text{int}}}a_i^{\text{int}},\\ \nonumber\label{EQM2}
\end{eqnarray}
where $G=g_{\text{nl}}| \alpha_p|^2$ describes the effective nonlinear coupling between idler and signal. These coupled equations can be solved
\begin{eqnarray}
 \textbf{a}=\textbf{M} . \textbf{a}_\textbf{in}^{\text{ext}}+\textbf{N} . \textbf{a}_\textbf{in}^{\text{int}},
  \label{EQM2}
\end{eqnarray}
where $\textbf{a}=[a_{s},a_{i}^\dagger]^\text{T}$, $\textbf{a}_\textbf{in}^\text{ext/int}=[a_{s}^{\text {ext/int}},(a_{i}^{\text {ext/int}})^\dagger]^\text{T}$ with $\text{T}$ stands for the transpose of the matrix, and

\begin{equation}
    \textbf{M}(\omega)=\begin{pmatrix}
\frac{\sqrt{\kappa_s^{\text{ext}}}\chi_i^*}{\chi_s\chi_i^*-G^2} & \frac{iG\sqrt{\kappa_i^{\text{ext}}}}{\chi_s\chi_i^*-G^2} \\
-\frac{iG\sqrt{\kappa_s^{\text{ext}}}}{\chi_s\chi_i^*-G^2}  & \frac{\sqrt{\kappa_i^{\text{ext}}}\chi_s}{\chi_s\chi_i^*-G^2}
\end{pmatrix},\\
    \textbf{N}(\omega)=\begin{pmatrix}
\frac{\sqrt{\kappa_s^{\text{int}}}\chi_i^*}{\chi_s\chi_i^*-G^2} & \frac{iG\sqrt{\kappa_i^{\text{int}}}}{\chi_s\chi_i^*-G^2} \\
-\frac{iG\sqrt{\kappa_s^{\text{int}}}}{\chi_s\chi_i^*-G^2}  & \frac{\sqrt{\kappa_i^{\text{int}}}\chi_s}{\chi_s\chi_i^*-G^2}
\end{pmatrix},
\end{equation}
where $\chi_j(\omega)=\big[-i(\omega-\omega_j)+\kappa_j/2\big]$ is the inverse of the field susceptibility with $\chi_j^*=\chi_j^*(-\omega)$. Since we are interested in the state of the output fields in the current system, we can rewrite the Eq. (\ref{EQM2}) in terms of the output fields using the input-output theory $a_{j,\text{in}}^{\text{ext/int}}=a_{j,\text{out}}^{\text{ext/int}}+\sqrt{\kappa_j^{\text{ext/int}}}a_j$ with $j=s,i$, resulting in

\begin{eqnarray}
\textbf{a}=\textbf{M} . \textbf{a}_\textbf{out}^{\text{ext}}+\textbf{N} . \textbf{a}_\textbf{out}^{\text{int}}+\textbf{B}. \textbf{a},
  \label{EQM4}
\end{eqnarray}
where 
\begin{equation}
    \textbf{B}=\begin{pmatrix}
\frac{\kappa_s\chi_i^*}{\chi_s\chi_i^*-G^2} & \frac{iG\kappa_i^{\text{ext}}}{\chi_s\chi_i^*-G^2} \\
-\frac{iG\kappa_s^{\text{ext}}}{\chi_s\chi_i^*-G^2}  & \frac{\kappa_i^{\text{ext}}\chi_s}{\chi_s\chi_i^*-G^2}
\end{pmatrix},
\end{equation}
Note that the final term in this equation introduces reliance on the higher order of the output field operators $O([\textbf{a}_\textbf{out}^{\text{ext/int}}]^n)$. However, for our analysis, we will disregard this term as our focus is solely on the generation of photon pairs.

The state of photons within the FDMR loop can be determined using the time-dependent Hamiltonian given by $|\psi\rangle=\mathcal{T}e^{-i/\hbar \int_{0}^t dtH(t)} |\psi (0)\rangle$, where $\mathcal{T}$ denotes the time order operator and $|\psi (0)\rangle=|0\rangle_{s, i}|0\rangle_{\mathrm{env}}$ represents the initial state of the idler, signal, and environment. Here, all modes are initially in the vacuum state, and hence, we have $a^\text{ext}_{s/i}|0\rangle_{s/i}=a^\text{int}_{\text{s/i}}|0\rangle_{env}=0$. For weak driving fields, the two-photon state can be simplified
\begin{eqnarray}
 |\psi\rangle \simeq\Big[1-\frac{i}{\hbar}\int_{0}^t dtH(t)\Big]  |\psi (0)\rangle=\Big[1-i\int_{0}^t dt\big(g_{\text{nl}} a_p^2 a_s^\dagger a_i^\dagger+h.c)\Big]  |\psi (0)\rangle,
 \label{psi}
\end{eqnarray}
where we only considered the nonlinear part of the Hamiltonian. 
Utilizing the frequency domain representation of the operators
\begin{equation}
    O(t)=\frac{1}{\sqrt{2\pi}}\int d\omega e^{-i\omega t}O(\omega),
\end{equation}
we get 
\begin{eqnarray}
 |\psi\rangle=\Big[1-\frac{ig_{\text{nl}}}{(2\pi)^{3/2}}\int_{0}^t dt \int d\omega d\omega'\omega'' \Big(a_p(\omega)^2 a_{s}^{\dagger}(\omega') a_{i}^{\dagger} (\omega'') e^{-i(2\omega-\omega'-\omega'')t}+h.c\Big)\Big]  |\psi (0)\rangle,
\end{eqnarray}
Replacing the pump drive with its classical form $a_p(\omega)\rightarrow |\alpha_p(\omega)|\delta(\omega_p)$ and perform the time integral, results in
\begin{eqnarray}
 |\psi\rangle&=\Big[1-\frac{ig_{\text{nl}}}{2\pi}\int d\omega' d\omega'' \Big(|\alpha_p(\omega_p)|^2 a_{s}^{\dagger}(\omega') a_{i}^{\dagger} (\omega'') +h.c\Big)\delta (2\omega_p-\omega'-\omega'')\Big]  |\psi (0)\rangle\\
 &=\Big[1-\frac{iG}{2\pi}\int d\omega \Big(a_{s}^{\dagger}(\omega) a_{i}^{\dagger} (2\omega_p-\omega) +h.c\Big)\Big]  |\psi (0)\rangle,
\end{eqnarray}
where the term $\delta (2\omega_p-\omega'-\omega'')$ fullfiles the phase matching condition $2\omega_p=\omega_s+\omega_i$ required for SFWM. The output state $|\psi\rangle_{\mathrm{out}}$ can be obtained by substituting $a_{s/i}$ with $a_{s/i}^{\text{out}}$ using Eq. (\ref{EQM2})
\begin{eqnarray} \label{state}
 |\psi\rangle_{\text{out}}&=\int \frac{d\omega}{2\pi} \Big[\zeta_{00} (\omega)  |0\rangle_{s,i} |0\rangle_{\text{env}}-\zeta_{11} (\omega)  |1_\omega,1_{2\omega_p-\omega}\rangle_{s,i} |0\rangle_{\text{env}}-\zeta_{10} (\omega)  |1_\omega,0\rangle_{s,i} |0,1_{2\omega_p-\omega}\rangle_{\text{env}}\\
 &-\zeta_{01} (\omega)  |0,1_{2\omega_p-\omega}\rangle_{s,i} |1_\omega,0\rangle_{\text{env}}-\eta(\omega)|0\rangle_{s,i} |1_\omega,1_{2\omega_p-\omega}\rangle_{\text{env}}\Big], \nonumber
\end{eqnarray}
with
\begin{eqnarray}
 \zeta_{00}(\omega)&=&2\pi \delta (\omega)-iG\big[M_{12}^*(\omega) M_{22}(2\omega_p-\omega)+N_{12}^*(\omega) N_{22}(2\omega_p-\omega)\nonumber\\&+&M_{11}(\omega) M_{21}^*(2\omega_p-\omega)+N_{11}(\omega) N_{21}^*(2\omega_p-\omega)\big],\nonumber\\
 \zeta_{11}(\omega)&=&iG\big[M_{11}^*(\omega) M_{22}(2\omega_p-\omega)+M_{12}(\omega) M_{21}^*(2\omega_p-\omega)\big],\nonumber \\
  \zeta_{10}(\omega)&=&iG\big[M_{11}^*(\omega) N_{22}(2\omega_p-\omega)+N_{12}(\omega) M_{21}^*(2\omega_p-\omega)\big],\\
    \zeta_{01}(\omega)&=&iG\big[N_{11}^*(\omega) M_{22}(2\omega_p-\omega)+M_{12}(\omega) N_{21}^*(2\omega_p-\omega)\big],\nonumber\\
  \eta(\omega)&=&iG\big[N_{11}^*(\omega) N_{22}(2\omega_p-\omega)+N_{12}(\omega) N_{21}^*(2\omega_p-\omega)\big],\nonumber  
\end{eqnarray}
In Eq. (\ref{state}), the initial term represents the presence of signal and idler in a vacuum state. The subsequent component characterizes the idler-signal two-photon states, giving rise to coincidences between the idler and signal modes. The third and fourth terms correspond to situations where one photon is present in either the idler or signal mode. This occurrence arises due to the loss of an idler or signal photon, and these terms contribute to the single count rates. Finally, the last term illustrates the loss of the generated photons to the surrounding environment. The parameters $|\zeta_{ij}(\omega)|^2$ and $|\eta(\omega)|^2$ quantify the probability of each of these described events. Based on this, the photon-pair generation rate is expressed as \cite{Grassani2015, 4wmixing_sipe, review_sfwm}

\begin{equation}
 N_{c}=\int_{-\infty}^{\infty}d\omega |\zeta_{11}(\omega)|^2=G^2 \int_{-\infty}^{\infty}d\omega \Big|M_{11}^*(\omega) M_{22}(2\omega_p-\omega)+M_{12}(\omega) M_{21}^*(2\omega_p-\omega)\Big|^2. 
\end{equation}
The second term of the above equation $M_{12} M_{21}^*\propto G^2$  represents a higher order pump-dependent correction of the coincidence. In the weak pumping regime $G^2\ll \kappa_s \kappa_i$ we can simplify the above equation
\begin{equation}
 N_{c}\approx G^2\int_{-\infty}^{\infty}d\omega |M_{11} (\omega)|^2 |M_{22} (2\omega_p-\omega)|^2\approx G^2 \int_{-\infty}^{\infty}d\omega \Big[\frac{\kappa_s^\text{ext}}{(\omega-\omega_s)^2+\kappa_s^2/4}\Big]\times \Big[\frac{\kappa_i^\text{ext}}{(2\omega_p-\omega-\omega_i)^2+\kappa_i^2/4}\Big]. 
\end{equation}
We now can perform the frequency integral and find an analytical expression for the coincidence rate
\begin{equation}
 N_{c}\approx  \frac{8\pi G^2(\kappa_s^\text{ext}\kappa_i^\text{ext})(\kappa_s+\kappa_i)}{\kappa_s\kappa_i\Big[(\kappa_s+\kappa_i)^2+(2\omega_p-\omega_s-\omega_i)\Big]}, 
\end{equation}
Considering the phase matching condition $2\omega_p=\omega_s+\omega_i$ we find
\begin{equation}
 N_{c}=  \frac{8\pi G^2(\kappa_s^\text{ext}\kappa_i^\text{ext})}{\kappa_s\kappa_i(\kappa_s+\kappa_i)}=\frac{8\pi g_{\text{nl}}^2P^2}{(\hbar \omega_p)^2(\kappa_s+\kappa_i)}\Big(\frac{\kappa_s^\text{ext}}{\kappa_s}\Big)\Big(\frac{\kappa_i^\text{ext}}{\kappa_i}\Big)\Big(\frac{\kappa_p^\text{ext}}{\kappa_p^2}\Big)^2, 
\end{equation}
In the symmetric coupling rates $\kappa=\kappa_s=\kappa_i=\kappa_p$ and critical coupling regime $\kappa_j^\text{ext}/\kappa_j=1/2$ the above equation reduces to
\begin{equation}
 N_{c}=  \frac{\pi g_{\text{nl}}^2P^2}{4(\hbar \omega_p)^2\kappa^3}, 
\end{equation}
as reported in the main manuscript. In reality, the measured coincidence rate deviates from this equation and is influenced by the detection efficiency and losses in the measurement chain, leading to the modified expression $N_{\text{tot},c}=\eta_i\eta_sN_c$.

The single-count rates for the idler and signal can be expressed as follows
\begin{eqnarray}
   N_{s}=\int_{-\infty}^{\infty}d\omega |\zeta_{10}(\omega)|^2&\approx\frac{8\pi g_{\text{nl}}^2P^2}{(\hbar \omega_p)^2(\kappa_s+\kappa_i)}\Big(\frac{\kappa_s^\text{ext}}{\kappa_s}\Big)\Big(\frac{\kappa_i^\text{int}}{\kappa_i}\Big)\Big(\frac{\kappa_p^\text{ext}}{\kappa_p^2}\Big)^2, \\
  N_{i}=\int_{-\infty}^{\infty}d\omega |\zeta_{01}(\omega)|^2&\approx \frac{8\pi g_{\text{nl}}^2P^2}{(\hbar \omega_p)^2(\kappa_s+\kappa_i)}\Big(\frac{\kappa_s^\text{int}}{\kappa_s}\Big)\Big(\frac{\kappa_i^\text{ext}}{\kappa_i}\Big)\Big(\frac{\kappa_p^\text{ext}}{\kappa_p^2}\Big)^2,  
\end{eqnarray}
Note that the leakage/residue of the pump, the efficiency of the measurement chain $\eta_j$, and the dark counts $D_j$ of the photodetectors add extra terms to the single-count rates
\begin{eqnarray}
   N_{\text{tot},j}=\eta_{j} N_{j}+N_{j,p}+D_{j},
\end{eqnarray}
with $j=s,i$. Note that the additional noise photons arising from the pump leakage $N_{j,p}=\beta_{j} P$ shows a linear dependency to the pump power $P$ \cite{Clemmen2009, Engin2013}. The coefficient $\beta_j$ is affected by imperfections in the bandpass filter and the presence of laser sidebands. These imperfections enable pump photons to reach the photodetectors, leading to an overall increase in the total single counts integrated over the measurement time.


\subsection{Device design and Fabrication}
We conducted device simulations using Lumerical software with the Mode Solution and Finite-Difference Time-Domain solver. Our simulations involved introducing a TE-polarized mode into waveguide couplers, where the coupling lengths matched the side length of the squared ring.
From the simulation results, we calculated the coupling angle $\theta$ and power coupling coefficient $p_c^2=\text{sin} (\theta)^2$. Subsequently, we designed the sample and fabricated the device on an SOI chip. The silicon waveguides within our TPI have dimensions of 450 nm in width and 220 nm in height, while the cladding consists of $2\mu$m of top SiO$_2$ and $2.2\mu$m of bottom SiO$_2$.
Our design includes square-shaped microrings with side lengths of 29.64 $\mu$m, and we integrated rounded corners with a radius of 5 $\mu$m to minimize scattering losses. To achieve efficient power coupling, we precisely set the gaps between the microrings at 180 nm, enabling approximately $98\%$ of power transfer from one microring to the next-nearest microring.
This particular design was chosen to attain the desired anomalous Floquet insulator behavior at a wavelength of 1545 nm.

\subsection{Activation of the FDMR}

We initiate the excitation of an FDMR and achieve precise tuning of its resonant frequency by manipulating the phase of one of the ring resonators situated on the lower boundary of the lattice. To facilitate this tuning process, we fabricate a titanium-tungsten (TiW) heater, which covers the rectangular perimeter, and apply current to the heater. As we perform this manipulation, we observe that the resonant wavelength shift $\delta \lambda$ of the FDMR exhibits an approximately linear relationship with the applied heater power $p_\text{heat}$, see Supplementary Information. An interesting feature that needs to be considered in designing the FSR of FDMRS, is that although the size of FDMRs is almost three times the size of a single resonator, the FSR of the FDMR is the same as the FSR of a single microring resonator. As we showed in Fig. 1c of the main text, by detuning the phase of a microring, up to three FDMRs can appear in topological bandgaps in one FSR of a single microring. Unlike other sorts of resonators in that, a single mode creates the resonance, each FDMR is the superposition of three identical Floquet modes in that each Floquet mode has the FSR of a single microring resonator. As seen in Fig. 1d of the main text, these three FDMRS are not identical, in terms of the Q-factor and the transmission dip, since they are made with different Floquet modes present in different bandgaps. To avoid any confusion between the FSR of FDMR and the frequency distance between excited FDMRs in one FSR, we defined the $\Delta_\text{FDMR}$. The method to find the exact value of $\Delta_\text{FDMR}$ has been previously discussed \cite{Afzal2021}.

\subsection{Measurement setup}
Figure 2a of the main text illustrates our experimental setup designed to test the correlation between the generated photon pairs and evaluate their quantum properties. The setup involves the following components: TE-polarized pump photons generated by an FPC and a tunable continuous-wave laser (Santec TSL-550) at telecom wavelengths. The laser output power ranges from 0.086 mW to 1.73 mW, delivered via a lensed-tip fiber that is butt-coupled to the chip's facet. The pump and generated wavelengths are collected at the output fiber, with $1\%$ of the light sent to a photon detector and a power meter to measure the device's transmission spectrum. The remaining light is split into two separate paths using a 50/50 coupler. The signal and idler photons then pass through bandpass filters (Optoplex C-Band 50GHz) to suppress the input pump light by 40dB, as well as other generated photons at different frequencies. Finally, two SNSPDs "ID Quantique - ID281" detect isolated photons, providing electrical signals to a Time Controller (ID900) with a timing resolution of 100ps. 

In our experiments, we initially inject 1mW of pump power from the laser into the lattice at the bottom left edge. We measure the output light from the bottom right edge while sweeping the laser wavelength from 1530nm to 1557nm (5 FSR of the TPI). The total loss from the laser to the power meter is measured as -10.5 dB, accounting for input/output coupling losses (from fiber to chip and vice versa), FPC, and loss in the chip. The losses in the FPC and the chip are measured as 0.24 dB and 2.6 dB/cm, respectively. As a result, for 1mW laser power, the power in the waveguide is estimated as 0.29 mW. The transmission spectrum before and after applying phase detuning is shown in Fig. 2c in main text, respectively by blue and pink traces.

\subsection{Single and coincidence counts measurement}

\begin{figure}[t]
    \centering
 \includegraphics [width=0.9\linewidth]{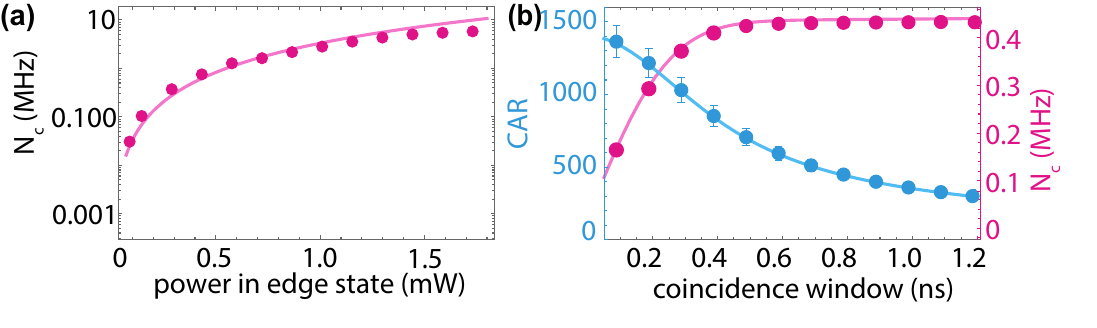}
    \caption{\textbf{a} Coincidence rate $N_c$ versus the pump power in the edge state at coincidence window of 300 ps. \textbf{b} Dependency of CAR and coincidence rate $N_c$ to the measurement coincidence window.}
    \label{Fig1methods}
\end{figure}

Taking into account the losses in both the signal and idler channels, the measured coincidence rate can be expressed as $N_{\text{tot},c}=\eta_i\eta_sN_c$, where $\eta_{s/i}$ represents the transmission efficiency from the chip's output to the SNSPDs, and $N_c$ denotes the inferred coincidence rate excluding measurement losses and efficiencies (see Supplementary Information). In addition to system losses, for the measured signal and idler pair rate, we must consider the noises and residues related to the sidebands of the pump and pump leakage, which exhibit a linear relationship with pump power ($N_{s/i,p} \propto P$). Furthermore, we account for the extra counts associated with the dark counts $D_{s/i}\approx 50$ Hz of the SNSPDs. Consequently, the measured single/idler counts rate can be defined as follows:
\begin{equation} 
    N_{\text{tot},s/i}= \eta_{s/i} N_{s/i}+N_{s/i,p}+D_{s/i}
    \label{Eq: Single_Count}
\end{equation}
where $N_{s/i}$ shows the inferred signal/idler single count rates. In Fig. \ref{Fig1methods}a, we plot the inferred $N_c$ as a function of the pump power in the edge state. As anticipated, the coincidence rate exhibits a rise with the increasing pump power.

\subsection{Coincidence-to-accidental rate}
 

The determination of the coincidence-to-accidental rate (CAR) involved using the coincidence histogram. We achieved this by calculating the ratio of the total coincidence counts, within a coincidence window centered around the peak values, to the accidental counts recorded during the same coincidence window \cite{Caspani2017}
\begin{equation}
    \text{CAR}=\frac{\int_{-t/2}^{t/2}g^2_{si}(t)dt}{\int_{\infty-t/2}^{\infty+t/2}g^2_{si}(t)dt}.
\end{equation}

To obtain the proper coincident window, we first fitted the coincidence counts using a Gaussian function to calculate the variance, $\delta=99.7$ ps and the full-width half maximum (FWHM) of the coincidence counts as $\text{FWHM}=2\sqrt{2\text{ln}2}\delta=234.8$ ps. 
Figure~\ref{Fig1methods}b shows the integration of coincidence count rates, $N_c$, over various coincidence windows and their corresponding CARs. As shown in this figure, there is a triad-off between the coincidence rate and CAR at different time windows. Since the coincidence rate gets saturated at coincidence windows greater than 0.5 ns, leading to dropping CAR, we consider the coincidence window$=3\delta\approx1.2\text{FWHM}=0.3$ ns to calculate CAR at different pump power shown in Fig. 3f of main text. We note the bandwidth of the generated photons (21 pm) is $75\%$ of the bandwidth of the FDMR (28 pm).
Given that the bandwidth of our tunable filters ($200$pm) exceeds the FDMR bandwidth but narrower than the edge state bandwidth to ($\approx 1$ nm), the filters cannot influence the measurements presented in Fig. 3c and f of main text.

\subsection{Power transmission spectra}

In this section, we analyze the transmission spectrum of the FDMR, considering the single-mode Kerr nonlinearity, as depicted in the effective resonator-waveguide representation shown in Fig. \ref{FigSI2}a. To start, we set $a_\text{res}=a_p=a_s=a_i$ in the Hamiltonian (\ref{Ham1}). This allows us to derive the Hamiltonian that describes the single-mode Kerr nonlinearity and its role in the transmission behavior of the nonlinear FDMR
\begin{equation}
    H= \hbar \omega_{\text{FDMR}}a_{\text{res}}^\dagger a_{\text{res}}-\hbar g_{\text{nl}} a_{\text{res}}^{\dagger 2} a_{\text{res}}^{2},
\end{equation}

The equation of motion describing the dynamics of the system then is given by
\begin{equation}
     \Dot{a}_\text{res}= -\Big(i\big[\omega_{\text{FDMR}}-2g_{\text{nl}}a_\text{res}^\dagger a_\text{res}\big] +\frac{\kappa}{2}\Big)a_\text{res}+\sqrt{\kappa^{\text{ext}}}a^{\text{ext}}+\sqrt{\kappa^{\text{int}}}a^{\text{int}},
     \label{Singlemode}
\end{equation}
where $\kappa=\kappa^{\text{ext}}+\kappa^{\text{int}}$, with $\kappa^{\text{ext}}$ and $\kappa^{\text{int}}$ represent the extrinsic and intrinsic coupling rates, respectively. Furthermore, $a^{\text{ext}}$ and $a^{\text{int}}$ are the annihilation operators of the input field from the edge state and loss channels, respectively. In the frequency domain, we can solve the coherent response of Eq. (\ref{Singlemode}), leading to
\begin{equation}
    \langle a_\text{res} (\omega)\rangle= \frac{\sqrt{\kappa^{\text{ext}}}\langle a^\text{ext}(\omega)\rangle}{\kappa/2 -i\Big ( \omega-\omega_{\text{FDMR}} +2g_{\text{nl}} \langle a_\text{res}^\dagger a_\text{res} \rangle \Big )},
    \label{transmission-res}
\end{equation}
here we consider $\langle a^{\text{int}} \rangle =0$. Using the input-output theory $a_\text{out}=a^\text{ext}-\sqrt{\kappa^{\text{ext}}}a_{\text{res}}$ we can find the coherent transmission of the system 
\begin{equation}
    T_0(\omega)=\frac{ \langle a_\text{out} (\omega)\rangle}{ \langle a_\text{ext} (\omega)\rangle}= 1- \frac{\kappa^{\text{ext}}}{\kappa/2 -i\Big ( \omega-\omega_{\text{FDMR}} +2g_{\text{nl}} n_\text{res}  \Big )}
    \label{transmission-res2}
\end{equation}
which has a nonlinear dependency to the photons inside the cavity $n_{\text{res}}=\langle a_\text{res}^\dagger a_\text{res} \rangle$, as expected for a cavity-based Kerr nonlinear platforms. In the low-power regime, the nonlinear term $2g_\text{nl}$ in the equation becomes negligible, and the transmission function $T_0$ follows the standard Lorentzian function of a cavity.

The Total transmission of the system is affected by several factors, which include the reflection and transmission at the input ($r_1, t_1$) and output ($r_2, t_2$) facets of the waveguide, as well as the coupling to and from the edge state. By taking these reflection and transmission parameters into account, we can determine the normalized power transmission at the output of the lensed coupled waveguide facet \cite{Gao2022}
\begin{equation}
    T(\omega)= \frac{ |t_1t_2|^{2} |T_0(\omega)|^2 }{ \Big| 1-r_1r_2 T_0(\omega)^2  e^{-i\phi} \Big|^{2} },
    \label{Trans}
\end{equation}
where $\phi$ represents the propagation phase of light in the coupling waveguide and edge state, which depends on the resonance frequency and propagation length. To fit the transmission spectrum of the FDMR, as shown in Fig. 2d of the main manuscript, we used Eq. (\ref{Trans}).

\subsection{Second-order correlation function for FDMR }
\begin{figure}[ht]
  \centering
  \includegraphics[width=0.95\textwidth]{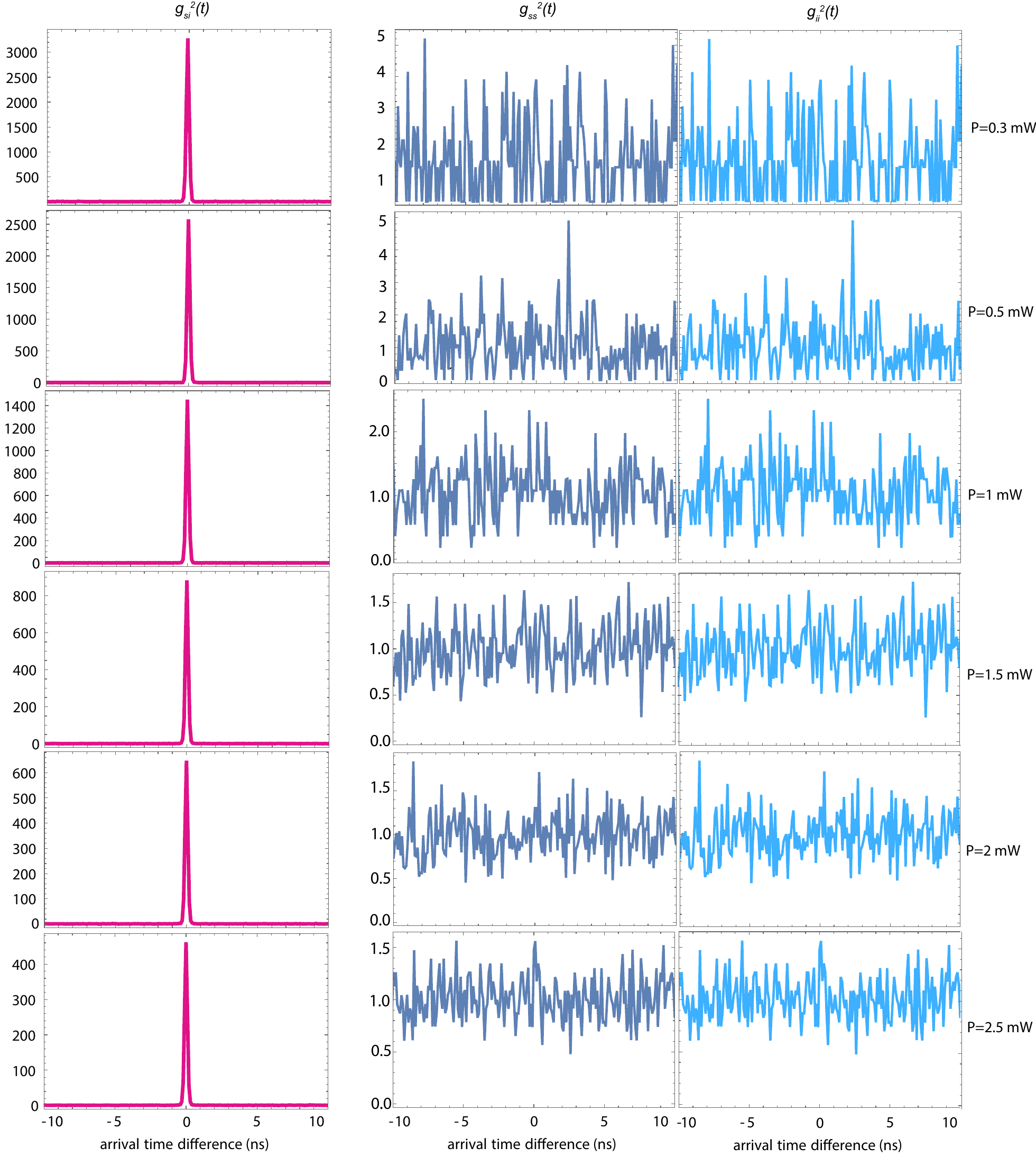}
  \caption{Comparing the second-order cross-correlation function $g^{(2)}_{si} (t)$, and auto-correlation functions of signal, $g^{(2)}_{ss}(t)$, and idler, $g^{(2)}_{ii}(t)$, for different pump powers at the presence of the FDMR.}
  \label{gsi_gss_gii}
  \end{figure}

Figure \ref{gsi_gss_gii} reports the raw data of the second-order cross-correlation function $g^{(2)}_{si}$ and auto-correlation functions of the idler and signal $g^{(2)}_{ss}$ and $g^{(2)}_{ii}$ for different pump powers at the presence of the FDMR. The result indicates the violation of the Cauchy-Schwarz inequality $[g^{(2)}_{si} (t)]^2 \leq g^{(2)}_{ss}(t)  \times g^{(2)}_{ii}(t) $ at arrival time difference $t=0$.  

\subsection{Coincidence rate: FDMR versus edge state}
In this section, we present the raw data of the measured coincidence counts of the FDMR and edge state, including the detection efficiency and setup loss without calibration i.e. $N_{\text{tot},c}=\eta_i\eta_sN_c$. Due to the small size of our sample and the low efficiency of photon conversion in SFWM, the edge state cannot generate significant photon pairs in the system. This limitation is attributed to the short interaction time available for generating pair-correlated photons in the edge state, which is considerably smaller compared to the resonance condition.

\begin{figure}[ht]
  \centering
  \includegraphics[width=0.8\textwidth]{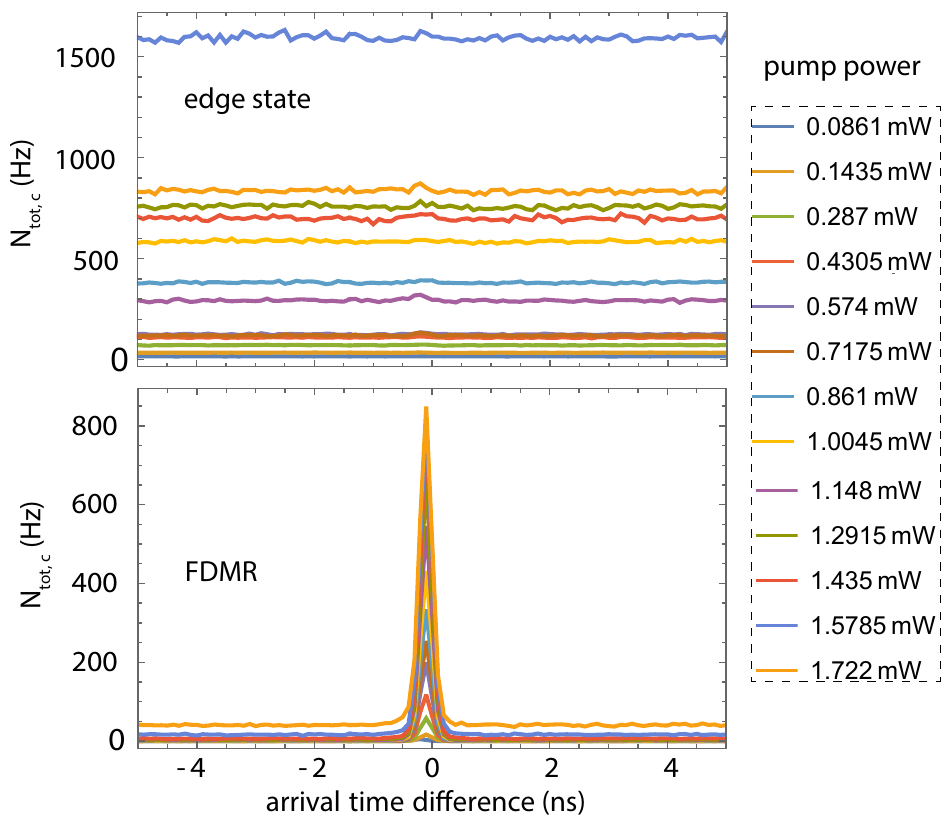}
  \caption{The measured (uncalibrated) coincidence rate of the edge state and the FDMR as a function of arrival time at different pump powers.  }
  \label{cc-time arrival}
\end{figure}

As we increase the power of the laser, noise and pump leakage to the photodetector become prominent. Therefore, no substantial coincidences have been observed in the edge state for various laser powers. In contrast, when the FDMR is activated, a significant level of coincidence has been detected. Figure \ref{cc-time arrival} illustrates that the noise level increases as the laser power is raised.

\subsection{Phase detune calibration}

\begin{figure}[ht]
  \centering
  \includegraphics[width=0.6\textwidth]{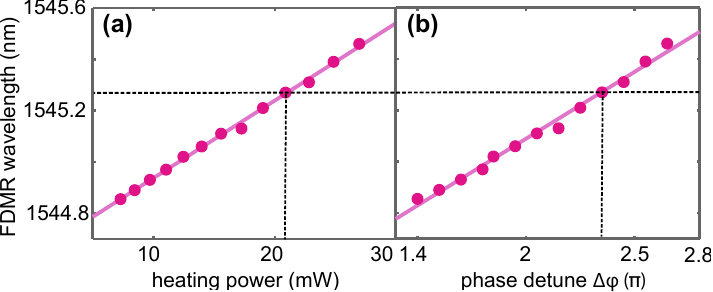}
  \caption{Phase detune calibration.  \textbf{a}, measured FDMR wavelength versus heating power, \textbf{a}, simulated FDMR wavelength versus applied phase detune over microring B at the bottom boundary of a microring lattice characterized with the parameters obtained from experimental measurements.}
  \label{Fig:FigSI5}
\end{figure}

In order to establish a calibration for the applied heating power in our experimental setup, and to ascertain its relationship with the introduced phase detuning, denoted as $\Delta\phi$, we conducted a systematic investigation. Our approach involved the variation of the heating power within the range of 7 mW to 27 mW, while concurrently measuring the wavelength associated with the excited FDMR, as illustrated in Fig.~\ref{Fig:FigSI5}a. Additionally, we conducted simulations to characterize the transmission spectrum of the lattice, specifically focusing on the wavelength of the FDMR for varying levels of phase detuning, ranging from $\Delta \phi=1.5\pi$ to $2.65\pi$. Notably, this phase detuning was uniformly applied to the entirety of microring B at the bottom boundary of the lattice, situated at the lattice boundary, as depicted in Fig.~\ref{Fig:FigSI5}b.  Through comparison between the simulation results and the experimental data, we obtained the $\Delta\phi=2.37\pi$ for excitation of the FDMR at the wavelength of our pump power.

\section{Details of experimental setup for entanglement measurement}
Photon pairs are sent through one input port of the interferometer, while a portion of the pump laser is attenuated using a variable optical attenuator (VOA) and sent to another port as a reference light for the phase stabilization system. The signal and idler photons interfere on a $50:50$ beam splitter, which has two output arms of different lengths. The long arm of the unbalanced Michelson interferometer contains a piezo-electric ring with fiber looped around it. At one of the output ports, a $1:99$ beam splitter directs 1 percent of the output for locking the interferometer.
The photodetector receives the locking signal and sends it to an analog input on the data acquisition (DAQ) card. The DAQ card then outputs an analog signal to the piezo controller, which transmits it to the piezo ring. The DAQ card employs a proportional, integral, and derivative (PID) control loop that dynamically regulates the feedback sent to the piezo controller. This process in addition to a temperature controller keeps the interferometer aligned at the target set point.
Using a PID control loop we dynamically regulate the feedback sent to the piezo controller and keep the interferometer phase stabilized. Two circulators are used to separate input and output lights. The signal and idler will be separated from the pump light using tunable band-pass filters and sent to SNSPDs.

\begin{figure}[ht]
  \centering
  \includegraphics[width=1\textwidth]{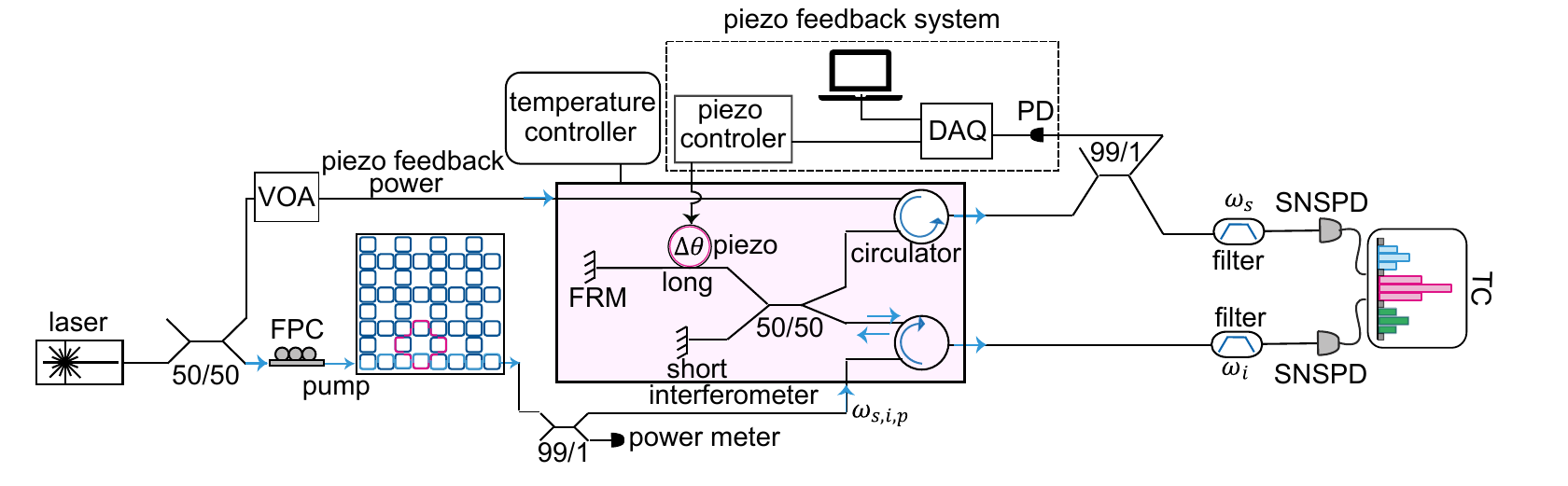}
  \caption{Measurement setup for Fiber-based unbalanced Michelson interferometer for energy-time entanglement measurement.}
  \label{Fig:FigSI_intanglement}
\end{figure}


\end{document}